\documentclass[article, nojss]{jss}

\usepackage{orcidlink,thumbpdf,lmodern}

\usepackage{subcaption}  

\usepackage{framed}
\usepackage{amssymb}
\usepackage{amsmath}
\usepackage{dsfont}

\newcommand{\dd}{\ensuremath{\,\textrm{d}}}

\DeclareMathOperator*{\argmin}{arg\,min}
\DeclareMathOperator*{\argmax}{arg\,max}

\newcommand{\norm}[1]{\left\Vert #1\right\Vert}

\newcommand{\one}{\mathds{1}}

\newcommand{\q}[1]{``#1''}

\usepackage[algoruled,linesnumbered,vlined]{algorithm2e}
\SetKwComment{Comment}{$\triangleright$\ }{}

\SetKwComment{MyComment}{}{}

\usepackage[textwidth=1.8cm, textsize=tiny]{todonotes}
\usepackage{booktabs}
\usepackage{multirow}

\usepackage{pifont}

\usepackage{tikz}
\usetikzlibrary{shapes, arrows.meta, positioning, trees}
\tikzstyle{folder} = [rectangle, rounded corners=5pt, draw=black, fill=yellow!30, minimum width=30pt, minimum height=15pt, text centered]

\usepackage{xcolor}
\definecolor{blue}{RGB}{78, 141, 196}
\definecolor{lightblue}{RGB}{135, 178, 215}
\definecolor{red}{RGB}{196, 78, 82}
\definecolor{lightred}{RGB}{215, 135, 138}
\definecolor{green}{RGB}{78, 196, 133}
\definecolor{brown}{RGB}{176, 138, 76}

\usepackage[edges]{forest}
\definecolor{folderbg}{RGB}{124,166,198}
\definecolor{folderborder}{RGB}{110,144,169}
\newlength\Size
\tikzset{%
  folder/.pic={%
    \filldraw [draw=folderborder, top color=folderbg!50, bottom color=folderbg] (-1.05*\Size,0.2\Size+5pt) rectangle ++(.75*\Size,-0.2\Size-5pt);
    \filldraw [draw=folderborder, top color=folderbg!50, bottom color=folderbg] (-1.15*\Size,-\Size) rectangle (1.15*\Size,\Size);
  },
  file/.pic={%
    \filldraw [draw=folderborder, top color=folderbg!5, bottom color=folderbg!10] (-\Size,.4*\Size+5pt) coordinate (a) |- (\Size,-1.2*\Size) coordinate (b) -- ++(0,1.6*\Size) coordinate (c) -- ++(-5pt,5pt) coordinate (d) -- cycle (d) |- (c) ;
  },
}
\forestset{%
  declare autowrapped toks={pic me}{},
  pic dir tree/.style={%
    for tree={%
      folder,
      font=\ttfamily,
      grow'=0,
    },
    before typesetting nodes={%
      for tree={%
        edge label+/.option={pic me},
      },
    },
  },
  pic me set/.code n args=2{%
    \forestset{%
      #1/.style={%
        inner xsep=2\Size,
        pic me={pic {#2}},
      }
    }
  },
  pic me set={directory}{folder},
  pic me set={file}{file},
}

\author{Romain E. Lacoste~\orcidlink{0009-0006-7029-8000}\\Université Gustave Eiffel}
\Plainauthor{Romain E. Lacoste}

\title{\pkg{Sparklen}: A Statistical Learning Toolkit for High-Dimensional Hawkes Processes in \proglang{Python}}
\Plaintitle{Sparklen: a statistical learning toolkit for high-dimensional Hawkes processes in Python}
\Shorttitle{\pkg{Sparklen}: toolkit for Hawkes Processes in \proglang{Python}}

\Abstract{
This paper introduces \pkg{Sparklen}, a statistical learning toolkit for Hawkes processes in \proglang{Python}, designed to bring together efficiency and ease of use. The purpose of this package is to provide the \proglang{Python} community with a complete suite of cutting-edge tools specifically tailored for the study of exponential Hawkes processes, with a particular focus on high-dimensional framework. It includes state-of-the-art estimation tools with built-in support for incorporating regularization techniques, and novel classification methods. To enhance computational performance, \pkg{Sparklen} leverages a high-performance \proglang{C++} core for intensive tasks. This dual-language approach makes \pkg{Sparklen} a powerful solution for computationally demanding real-world applications. Here, we present its implementation framework and provide illustrative examples, demonstrating its capabilities and practical usage.
}

\Plainkeywords{Hawkes processes, high-dimension, exponential kernel, empirical risk minimization, maximum likelihood estimation, regularization, optimization, Python, C++}

\Address{
  Romain E. Lacoste\\
  Université Gustave Eiffel\\
  5 Boulevard Descartes, 77420, Champs-sur-Marne, France\\
  E-mail: \email{romain.lacoste@polytechnique.edu}\\
  URL: \url{https://romain-e-lacoste.github.io/}
}

\begin{document}

\section[Introduction: Hawkes processes modeling in Python]{Introduction: Hawkes process modeling in \proglang{Python}} \label{sec:intro}

This paper introduce the \proglang{Python} package \pkg{Sparklen} (see \cite{Lacoste:2025}), which implements a complete set of statistical learning methods for exponential Hawkes processes with an emphasize on high-dimension setting. Hawkes processes, introduced in \cite{Hawkes:1971}, form a specific but rather versatile class of point processes. Such processes model time series in which the occurrence of one event temporarily increases the probability of other events occurring. This intrinsic ability to take into account self-exciting effects makes them particularly interesting for real data modeling. Historically applied in seismology (see \cite{Ogata:1988}), they have since been used in a wide variety of other fields, including neuroscience in \cite{Reynaud-Bouret:2013}, finance in \cite{Bacry:2015}, ecology in \cite{Denis:2024}. The multidimensional version, known as the Multivariate Hawkes Processes (MHP), captures additionally interactions among each univariate process within a network. This generalization enables the modeling of more intricate dynamics, significantly expanding the range of potential applications. For example, MHP has been applied to model action potentials within neural networks in \cite{Bonnet:2022}, or for trend detection in social networks in \cite{Pinto:2015}. 

\paragraph{Software for MHP.} To support this virtuous trend of applications, considerable effort has gone into the development of statistical methods for MHP. Bridging the gap between theoretical advances and practical applications necessitates the development of software that provides comprehensive and user-friendly tools. Over the years, packages for MHP have emerged, aiming to fulfill this need. A key reference in this field is \pkg{tick} (\cite{Bacry:2018}), a \proglang{Python} package that provides a list of tools for statistical learning with an emphasis on time-dependent models. Although \pkg{tick} offers a complete set of tools for MHP, it faces maintenance issues and is not installable in many recent \proglang{Python} configurations. Other packages, covering more specific uses and therefore being less comprehensive are also available. In the Bayesian framework, one can cite the \pkg{pyhawkes} (\cite{Linderman:2017}) \proglang{Python} package which proposes inference algorithms for discovering latent network structure given point process observations. In situations where count data are only observed in discrete time, the \pkg{hawkesbow} (\cite {Cheysson:2024}) \proglang{R} package provides an implementation of an estimation method for univariate Hawkes processes using a spectral approach derived from the Bartlett spectrum. Recently, the \pkg{FaDIn} package introduces a novel discretized inference procedure for MHP with parametric kernels of finite support (see \cite{Staerman:2023}). So, despite the growing popularity of MHP, the community lacks a comprehensive, accessible implementation environment that unifies the wide range of tools developed for theses processes. 

\paragraph{\pkg{Sparklen} software.} In this context, we introduce the \proglang{Python} package \pkg{Sparklen} (see \cite{Lacoste:2025}). The aim of this package is to provide to the \proglang{Python} community a complete set of easy to use and efficient statistical tools for MHP, with an emphasize on high-dimension setting. Before delving into the package's features, let us first examine its implementation architecture. From a syntactic standpoint, the code rigorously adheres to object-oriented programming principles and conforms to the widely used \pkg{scikit-learn} API (see \cite{Pedregosa:2011}; \cite{Buitinck:2013}). This ensures a familiar and efficient framework for users, simplifying and streamlining their workflow. The package's structure is thoughtfully organized, with each functionality carefully compartmentalized. A key strength lies in its use of core \proglang{C++} code for computationally intensive components, ensuring both efficiency and performance. The binding between \proglang{C++} and \proglang{Python} is handled through \pkg{SWIG} wrapper code (see \cite{Beazley:2003}). In essence, \pkg{SWIG} facilitates interoperability by generating \proglang{Python} code that wraps \proglang{C++} classes through written interface files. In a nutshell, this dual-language approach combines the accessibility of \proglang{Python} with the computational efficiency of \proglang{C++}. This makes \pkg{Sparklen} highly user-friendly, while also ensuring it is well-suited for computationally demanding applications. Finally, this package is thoroughly documented, further enhancing its accessibility.

\paragraph{Functionality and scope.} The \pkg{Sparklen} library offers a collection of modules for optimization with regularization, featuring classes for optimizers, learning rate schedulers, proximal operators, and calibration methods. These components can be flexibly combined and integrated to meet various needs. The core module of the package is dedicated to exponential MHP. The exponential kernel is a popular choice due to its computational efficiency and its structure, which facilitates straightforward interpretability. At this stage, we wish to highlight that all the tools implemented in this package are equally applicable to univariate Hawkes processes, as these are naturally included as a special case of the MHP model. The \pkg{Sparklen} package integrates several tools for working with MHP, such as an efficient cluster-based simulation method for generating MHP events. The package's main features are categorized into two key areas: inference and multiclass classification methods for MHP, with a particular emphasis on the high-dimensional framework.

\paragraph{Inference in \pkg{Sparklen}.}
First, let us outline the inference tools provided in the \pkg{Sparklen} package. In the literature of MHP, particular attention has been paid to the design of efficient estimating procedures. For example, maximum likelihood approaches have been proposed, one can cite \cite{Ogata:1978}. In addition, methods relying on least square contrast exist in the literature (\cite{Reynaud-Bouret:2014}; \cite{Bacry:2020}). In this article, we consider an original setting, assuming that the data we have access to consist of repeated short-time paths. 
For this observational setup, one can cite the work of \cite{Lotz:2024} for likelihood approach and \cite{Dion-Blanc:2024} for least-squares. The \pkg{Sparklen} package implements both likelihood and least-squares approaches, handling the aforementioned setup. A key consideration is that the loss, gradient, and hessian of both functionals are computed exactly for precision and efficiently for speed. Furthermore, several regularizations are implemented, allowing users to incorporate prior knowledge about the nature of the problem. Notably, the tuning of the regularization constant, optimally selected internally based on various user-selectable criteria, is a very appealing contribution to this package. This makes the implemented tools highly versatile in a variety of situations. Specifically, the procedures are tailored for high-dimensional settings. In particular, the use of Lasso regularization combined with automatic tuning of the penalty constant can lead to successful recovery of the support of a high-dimensional MHP.

\paragraph{Classification in \pkg{Sparklen}.} Second, let us describe the multiclass classification setting we consider and the tools implemented in the \pkg{Sparklen} package to address such problem. Time series classification is a fast-growing and flourishing field of research, the fruit of a vigorous research effort in recent years (see \cite{Ismail:2019}). Due to their complex nature, standard learning algorithms are generally not tailored for working on raw time series. On the other hand, the problem of classifying time series, which consists in labeling data given that each class is characterized by a specific underlying temporal dynamic, naturally arises in real life. Examples of applications are legion, for example for ECG pattern recognition in \cite{Chaovalitwongse:2006}, for fault detection and cyber-security applications in \cite{Susto:2018}. As a result, there is a strong need to develop cutting-edge algorithms to process such data. Recently, some work has focused on the classification of Hawkes processes paths. For example, work in natural language for rumour stance classification can be found in \cite{Lukasik:2016}. 
In our context, we assume that the learning sample consists of \textit{i.i.d.} labeled data where the features are the jump times of a
MHP observed on the fixed time interval. In such a setting, an application can be found in ecology for modeling echolocation calls of bats to classify their behavior in several sites throughout France in \cite{Denis:2024}. The used classification procedure follows the strategy presented in \cite{Denis:2022} in the univariate case and have been recently extended to the multivariate case in \cite{Dion-Blanc:2024}. This novel procedure takes advantage of an initial interaction recovery step, by class, followed by a
refitting step based on a suitable classification criterion. The \pkg{Sparklen} package implements the two classification procedures. While both rely on empirical risk minimization, the second method, which includes a prior recovery step using a Lasso-type estimator, enables handling of high-dimensional settings.

\paragraph{Outline of the paper.} The remainder of this paper is structured as follows. In Section~\ref{sec:theory_hawkes}, we provide an overview of the theoretical foundations underpinning \pkg{Sparklen}'s features. Section~\ref{sec:implementation} presents the implementation and provides insightful examples that highlight \pkg{Sparklen}'s capabilities. Then in Section~\ref{sec:illustrations}, the package is illustrated on MemeTracker data set. Finally, in Section~\ref{sec:conclusion}, we conclude the paper and outline potential future developments for the package.

\section{Hawkes processes} \label{sec:theory_hawkes}

First, in Section~\ref{subsec:model}, we introduce the Hawkes model. Then, Section~\ref{subsec:simu_branch} presents an efficient simulation algorithm that leverages its branching structure. Section~\ref{subsec:inference} covers the inference tools available for estimating a MHP. Finally, in Section~\ref{subsec:classification}, we present the supervised classification framework and classification algorithms tailored to MHP paths. In the following, the set $\{1, \dots, n\}$ is denoted by $[n]$ for any integer $n \geq 1$.

\subsection{Hawkes model}
\label{subsec:model}

In many fields of application, data takes the form of points distributed in a space. For instance, it is usual to observe points distributed along the real line, such as the occurrence of events over time. A pivotal issue, for modeling and prediction purpose, is to provide a suitable probabilistic framework to describe the temporal distribution of these events. Point processes, with their associated formalism, constitute a powerful and flexible tool for modeling such distributions. 

A point process, denoted $N$, is encoded by its events arrival times $(t_{\ell})_{\ell \geq 1}$, given by the position of the points in $\mathbb{R}_+^*$. In addition, it can be described by its counting process, denoted $\left(N(t)\right)_{t \geq 0}$ and defined by $N(t) = \sum_{\ell \geq 1}\one_{\{t_{\ell} \leq t\}}$, which gives the number
of events that have occurred before time $t$. In the paradigm of point processes, a key component is the intensity function, which characterizes the process by dictating the dynamics of event occurrences. Such function, for a given $t \geq 0$, takes the following form:
$$\lambda(t):=\lim _{h \rightarrow 0^{+}} \frac{\mathbb{P}\left(N(t+h)-N(t)=1 \mid \mathcal{F}_{t^-}\right)}{h} $$
where the filtration $\mathcal{F}_{t^-}$ represents the information available strictly before time $t$. Heuristically, at a given time, it may be perceived as the infinitesimal probability of observing an event in the near future, conditionally on the past of the process. 

A classical example of point processes is given by the homogeneous Poisson process, which models the occurrence of events at rate $\lambda(t) = \mu$. In this case, the intensity is not random and does not depend on $t$. In particular, successive events are independent of one another. Despite being a quite popular tool, this feature of independence is not always realistic from a modeling standpoint. Indeed, in many applications, the observed dynamic induces more tricky temporal dependencies that cannot be captured by such a model. On the other hand, Hawkes processes are tailored to model time series for which the occurrence of events depends on past observations. 

\begin{leftbar}
This work centers on the linear Hawkes process with an exponential kernel, a well-regarded choice due to its favorable properties. Notably, the recursive formulation of the likelihood enables efficient computation, while the model’s structure further supports straightforward interpretability, enhancing its usability in practice.
\end{leftbar}

The multivariate version of these processes, called Multivariate Hawkes Processes (MHP), stands out as a natural extension enabling to describe more complex dynamics, thus broadening the spectrum of applications. In such a model, the evolution of a component's dynamics is no longer only likely to be impacted by the presence of self-exciting effects, but also by exciting interactions of components connected to it within the network. A $d$-MHP, where $d\geq1$ is the network's dimension, is denoted as $N = (N_1, \dots, N_d)$. Each component $N_j$ is characterized by its intensity function, defined for each $t \geq 0$ as follows:
\begin{equation}\label{eq:lambda_j}
    \lambda_j(t) := \mu_j + \sum_{j'=1}^d \alpha_{j,j'} \int_0^t \beta e^{-\beta(t-s)} \dd N_{j'}(s) = \mu_j + \sum_{j'=1}^d  \alpha_{j,j'} \sum_{\ell : t_{j', \ell} < t} \beta e^{-\beta(t-t_{j',\ell})}
\end{equation}
where $\mu = (\mu_j)_{j \in [d]}$ is the vector of exogenous intensities, $\alpha = (\alpha_{j, j'})_{j, j' \in [d]}$ is the matrix of interactions, and $\beta$ is the decay. For each $j \in [d]$, the coefficient $\mu_j > 0$ expresses the arrival of spontaneous events for the $j$-th process. For each $j, j' \in [d]$, the coefficient $\alpha_{j,j'}$ is non-negative and reflects the positive influence of $N_{j'}$ on $N_j$. Finally, the scalar $\beta > 0$ dictates how quick these influences vanish over time. 

\subsection{Simulation and branching structure} 
\label{subsec:simu_branch}

An appealing property of Hawkes processes is their branching structure. In particular, they can be represented as Poisson cluster processes, which are age-dependent immigration-birth processes. Consequently, each cluster forms a continuous-time Galton–Watson tree with a Poisson offspring process. Introduced in \cite{Hawkes:1974} for the univariate case, this bridge between branching theory and Hawkes processes enables to derive theoretical results of interest such as existence, asymptotic behavior as well as counting and interval properties of these processes. 
In addition to offering  a rich theoretical framework, this cluster representation gives rise to an efficient simulation procedure (presented in \cite{Møller:2005}; \cite{Chen:2020}). Indeed, by exploiting the hierarchical structure of descendants, one can simulate the exact number of events without any rejection, in contrast to the classical thinning approach (see \cite{Ogata:1981}).

Consider a $d$-MHP and let $T>0$ the (non-necessarily finite) horizon time. The simulation of the events on $[0, T]$ for this MHP is as follows. First, for each $j \in [d]$, we simulate the immigrants of the $j$-th component according to a homogeneous Poisson process with intensity $\lambda = \mu_j$ on $[0,T]$. These events are denoted $\{t_{j,\ell}^0\}_{\ell\in [m_j^0]}$ with $m_j^0 \sim \mathcal{P}(\mu_j T)$. These immigrants become the ancestors and we must simulate their offspring. For each $j$ and for each $\ell\in [m_j^0]$, consider the $\ell$-th ancestor of type $j$ at generation $0$. The latter, denoted $t_{j,\ell}^0$, generates descendants of generation $1$ in each dimension $j' \in [d]$ according to a inhomogeneous Poisson process with intensity function $\lambda_{j,j'}(\cdot- t_{j,\ell}^0) = \alpha_{j,j'}\beta \exp(-\beta(\cdot - t_{j,\ell}^0))$ on $[t_{j,\ell}^0, T]$. For $j \in [d]$, the events are denoted $\{t_{j,\ell}^1\}_{\ell\in [m_j^1]}$ with $m_j^1 \sim \mathcal{P}(\alpha_{j,j'}(1-\exp(-\beta(T-t_{j,\ell}^0)))$. Then, in the next step, these descendants become the ancestors and the descendants of generation 2 can be simulated, and so on. We continue iterating until there are no ancestors left in each dimension. This construction is precisely detailed in Algorithm~\ref{alg:sim_cluster}. 

\begin{algorithm}[H]
\SetAlgoLined
\SetKwInput{Input}{Input}
\SetKwInput{Init}{Initialization}
\SetKwInput{Output}{Output}
\SetKwInput{Step1}{Output}
\Input{$\mu \in \mathbb{R}_+^d$, $\alpha \in \mathbb{R}_+^{d \times d}$, $\beta>0$ and $T>0$} 
\Init{$\mathcal{T} = \{\mathcal{T}_j \mid \mathcal{T}_j=\emptyset, \ j \in [d]\}$, $\mathcal{A} = \{\mathcal{A}_j \mid \mathcal{A}_j=\emptyset, \ j \in [d]\}$}

\MyComment{Immigrant simulation}

\For{$j=1, \dots, d$}{
    $k \sim \mathcal{P}(\mu_j T)$ \Comment*[r]{Number of immigrants of type $j$}
    $t_{j,1}, \dots, t_{j,k} \overset{\text{i.i.d.}}{\sim} \mathcal{U}([0, T])$ \;
    $\mathcal{A}_j = \{t_{j,1}, \dots, t_{j,k}\}$ \;
    $\mathcal{T}_j = \mathcal{T}_j \cup \mathcal{A}_j$ \;
}

\MyComment{Offspring simulation}

\While{there exists at least one $j$ such that $\mathcal{A}_j \neq \emptyset$}{
    $\mathcal{O} = \{\mathcal{O}_j \mid \mathcal{O}_j=\emptyset, \ j \in [d]\}$ \Comment*[r]{Offspring initialization}
    \For{$j=1, \dots, d$}{
        \If{$\mathcal{A}_j \neq \emptyset$}{
            \For{each $a_{j,\ell} \in \mathcal{A}_j$}{
                \For{$j'=1, \dots, d$}{
                    \If{$\alpha_{j',j} > 0$}{
                        $k \sim \mathcal{P}(\alpha_{j',j})$ \;
                        $t_{j',1}, \dots, t_{j',k} \overset{\text{i.i.d.}}{\sim} \mathcal{E}(\beta)$ \;
                        $\mathcal{O}_{j'} = \mathcal{O}_{j'} \cup \{a_{j,\ell} + t_{j',1}, \dots, a_{j,\ell} + t_{j',k}\}$ \;
                    }
                    $\mathcal{T}_{j'} = \mathcal{T}_{j'} \cup \mathcal{O}_{j'}$ \;
                }
            }
        }
    }    
    $\mathcal{A} = \mathcal{O}$ \Comment*[r]{Offspring become ancestors}
}

\MyComment{Remove events beyond $T$ and sort}

\For{$j=1, \dots, d$}{
    $\mathcal{T}_j = \text{sort}(\{t_{j,\ell} \mid t_{j,\ell} < T\})$ \;
}
\Output{$\mathcal{T} = \{\mathcal{T}_j, \ j \in [d] \}$}
\caption{Simulation by cluster representation}\label{alg:sim_cluster}
\end{algorithm}

As outlined in Algorithm~\ref{alg:sim_cluster}, the descendants are simulated according to an inhomogeneous Poisson process on $[a_{j,\ell}, +\infty[$, followed by the removal of events exceeding $T$. By doing so, the number of events simplifies into $\alpha_{j, j'}$ as can be seen in line~$14$ in Algorithm~\ref{alg:sim_cluster}. This prevents from calculating the integral for every ancestor $a_{j,\ell}$ and each coefficient $\alpha_{j, j'}$. This trick helps accelerating the sampling procedure in practice. 

\subsection{Parametric inference methods}\label{subsec:inference}

Before presenting the parametric estimation strategies, let us first define the statistical framework we consider. As it was implied in Equation~\eqref{eq:lambda_j}, we suppose that the intensity function of the process $N$ depends on an unknown parameter $\theta^*$. It is therefore through this unknown parameter that the global dynamic of the process is encoded. In contrast, the parameter $\beta$ is assumed to be known for optimization purposes (see below). The unknown parameter of interest, $\theta^* = (\mu^*, \alpha^*)$, is supposed to belong to the following parameter family:
$$\Theta := \left\{\mu \in \left(\mathbb{R}_+^*\right)^d, \ \alpha \in \mathbb{R}_+^{d \times d}, \ \rho(\alpha) < 1 \right\} \subseteq  \mathbb{R}_+^{d \times d+1}$$
with $\rho(\alpha)$ the spectral radius of the matrix $\alpha$. This hypothesis, known as sub-critical stability hypothesis, is crucial to ensure the non-explosion of the process.

Let $T >0$ be fixed. We denote $\mathcal{T}_T := \left\{\{t_{j,\ell}\}_{\ell \in [N_j(T)]}, \ j \in [d] \right\}$ the event times over $[0,T]$ of a MHP $N=(N_1, \dots, N_d)$ of intensity $(\lambda_{1, \theta^*}, \dots, \lambda_{d, \theta^*})$. We assume that we have access to a $n$-sample $\mathcal{D}_n := \left\{\mathcal{T}_T^{(1)}, \dots, \mathcal{T}_T^{(n)}\right\}$ which consists of independent copies of $\mathcal{T}_T$. In other words, we are dealing with short-time path repetitions. In particular, we do not assume that observed paths have reached the stationary regime.  Given the observation of the sample $\mathcal{D}_n$, our strategy is to minimize the following optimization problem:

\begin{equation}\label{eq:optim}
    \hat{\theta}_n(\kappa) \in \argmin_{\theta \in \mathbb{R}^{d \times d+1}} \left\{F_{T, n}(\theta) + \kappa \Omega(\theta) \right\}
\end{equation}

where $F_{T, n}(\cdot)$ is a loss function describing how well the model fits the data, $\Omega(\cdot)$ a regularization function that penalizes complex solutions and $\kappa > 0$ the regularization constant which controls the level
of inductive bias. 

\begin{leftbar}
The minimization problem given by Equation~\eqref{eq:optim} is the cornerstone of this section. In what follows, we present an overview of the various options available for loss functions, regularization functions, penalty constants choice, and optimization methods. Each option is examined in light of current advancements in the field and its relevance in the context of the application concerned.
\end{leftbar}

\paragraph{Loss function.}
We focus on two parametric estimation approaches, distinguished by the choice of the following two loss functions:
\begin{itemize}
    \item Log-likelihood functional, $F_{T,n}(\theta) = L_{T,n}(\theta)$,
    \item Least-squares contrast, $F_{T,n}(\theta) = R_{T,n}(\theta)$.
\end{itemize}

The first, of more classical inspiration, is given by the maximum likelihood estimation method. Introduced in \cite{Ogata:1978}, the Maximum Likelihood Estimator (MLE) is shown to be consistent, asymptotically normal and asymptotically efficient when the time horizon $T$ tends to infinity for a stationary Hawkes process. In the setting we consider, namely that of repeated observations with $T$ fixed, the MLE is shown to be consistent as the number of observations $n$ tends to infinity (see \cite{Lotz:2024}). From a practical point of view, we rather consider the negative log-likelihood in order to use first-order optimization algorithms to find its minimizer (see below). The negative log-likelihood over the repetitions is given by:
$$L_{T,n}(\theta) := \sum_{i=1}^n \left(\sum_{j=1}^d \int_0^T \lambda_{j, \theta}^{(i)}(t) \dd t - \sum_{j=1}^d\sum_{\ell : t_{j, \ell}^{(i)} < T} \log\left(\lambda_{j, \theta}^{(i)}\left(t_{j, \ell}^{(i)}\right)\right)\right)$$

where $\lambda_{j, \theta}^{(i)}(\cdot)$ is defined from the observation $\mathcal{T}_T^{(i)}$. An alternative approach, receiving comparatively less attention, is founded on the least-squares contrast, drawing inspiration from the theory of empirical risk minimization (\cite{Reynaud-Bouret:2014}; \cite{Bacry:2020}). Such a functional benefits from attractive computational and optimization-related properties (as discussed below).
Recently, in \cite{Dion-Blanc:2024}, within our observational setup of repeated short-time paths, the authors propose a Lasso-type procedure based on this loss to handle high-dimensional issues for MHP. Notably, a result of uniform consistency of the Lasso estimator as well as the consistency of the estimated support is proven when $n$ tends to infinity.  
The least-squares contrast averaged over the repetitions takes the following form:
$$R_{T,n}(\theta) := \frac{1}{n} \sum_{i=1}^n \left(\frac{1}{T} \sum_{j=1}^d \int_0^T \lambda_{j, \theta}^{(i)2}(t) \dd t - \frac{2}{T} \sum_{j=1}^d\sum_{\ell : t_{j, \ell}^{(i)} < T}\lambda_{j, \theta}^{(i)}\left(t_{j, \ell}^{(i)}\right)\right).$$

For both functions, it is possible, by simplification, to isolate the terms that depend exclusively on the exponential function and the event arrival times. As these terms are independent of the parameter, they can be calculated once and reused thereafter. This calculation trick speeds up future evaluations of the loss, gradient or Hessian. The implementation adopts this approach for optimal efficiency. For more details, see Appendix~\ref{app:precompute}.

In addition, it is worth noting that the inclusion of $\beta$ among the parameters to be estimated would result in the loss of convexity for both functionals, thereby complicating the optimization process. As a result, the decision to fix $\beta$, as outlined previously, is a deliberate strategy aimed to circumvent such difficulties. In situations where $\beta$ is unknown, the practitioner may explore a grid of candidate values and select the most suitable $\beta$ based on an \textit{ad hoc} criterion. In Table~\ref{tab:comparison_loss}, a comparison of the two functions properties is provided. 

\begin{table}[ht]
    \centering
    \begin{tabular}{lccc|cc}
        \toprule
        \multirow{2}{*}{\textbf{Function}} & \multicolumn{3}{c|}{\textbf{Complexity}} & \multicolumn{2}{c}{\textbf{Property}} \\ 
        & \textbf{Pre-calculation} & \textbf{Loss} & \textbf{Gradient} & \textbf{Convex} & \textbf{$L$-Smooth} \\
        \midrule
        Log-likelihood & $\mathcal{O}(md)$ & $\mathcal{O}(md)$ & $\mathcal{O}(md)$ & \ding{51} & \ding{55}  \\
        Least-squares & $\mathcal{O}(md)$ & $\mathcal{O}(d^3)$ & $\mathcal{O}(d^3)$ & \ding{51} & \ding{51} \\
        \bottomrule
    \end{tabular}
    \caption{Comparison of the log-likelihood and least-squares functions in terms of algorithmic complexity and optimization properties. The total number of observed events is denoted by $m$. Note that for least-squares, once pre-calculations have been performed, the complexity of both the loss function and its gradient evaluation remains independent of $m$.}
    \label{tab:comparison_loss}
\end{table}

\paragraph{Regularization.}
In machine learning, regularization encompasses a set of techniques that emerge naturally to prevent the effects of over-fitting and increase generalization ability (see \cite{Bach:2024} for more details). The incorporation of a regularization term, inducing constrained optimization, reflects a compromise between minimizing the objective function and maintaining the structural simplicity of the model through the hyperparameter $\kappa$. The choice of the regularization function $\Omega(\cdot)$ may be dictated by a prior knowledge of the structure of the parameter $\theta^*$. In this paper, we consider three types of regularization: 
\begin{itemize}
    \item Lasso regularization by adding a $\ell_1$-penalty $\Omega(\cdot) = \norm{\cdot}_1$ term (see \cite{Tibshirani:1996})
    \item Ridge (or Tikhonov) regularization by incorporating a $\ell_2$-penalty $\Omega(\cdot) = \norm{\cdot}_2^2$ (see \cite{Hoerl:1970}) 
    \item Elastic-Net regularization by considering a linear combination of $\ell_1$-penalty and $\ell_2$-penalty  $\Omega(\cdot) = \zeta\norm{\cdot}_1 + (1-\zeta)\norm{\cdot}_2^2$ (see \cite{Zou:2005})
\end{itemize}

Each of these have its own typical use case. For instance, in the case of a large network with few effective interactions, one may use Lasso regularization for its intrinsic virtue of variable selection and improved interpretability.

\paragraph{Calibration of $\kappa$.} The penalization constant $\kappa$ controls the power of the induced regularization. Small values lead to
complex models that are likely to over fit, while on the other hand large values lead to simplistic models with poor prediction power. The question of choosing this value, which encapsulates the trade-off described before, is therefore decisive for the success of the learning phase. In this article, we examine two methodologies for making this choice: 
\begin{itemize}
    \item Cross validation (\textsc{CV}) (\textit{e.g.} \cite{Arlot:2010})
    \item Extended Bayesian Information Criterion (\textsc{EBIC}) (\textit{e.g.} \cite{Chen:2008})
\end{itemize}
For both methods, the strategy involves exploring a grid $\Delta$ of $\kappa$ values and selecting the one that minimizes the criterion of interest. In the case of CV, this entails partitioning the dataset into folds and aiming to minimize the cross-validated risk. In the case of \textsc{EBIC}, given $\gamma \in [0, 1]$, the criterion takes the following form:
$$\text{EBIC}_{\gamma}(\kappa) := -2 L_{T, n}\left(\hat{\theta}_n(\kappa)\right) + \left\lvert S_{\hat{\theta}_n(\kappa)} \right\rvert \log(n) + 2\gamma\log\left(\binom{M^2}{\left\lvert S_{\hat{\theta}_n(\kappa)} \right\rvert}\right)$$
where $\kappa \in \Delta$ is the penalization constant under consideration; $\hat{\theta}_n({\kappa})$ is the minimizer of the problem given by Equation~\eqref{eq:optim} (with any function $F_{T, n}(\cdot)$ and $\Omega(\cdot)$); $L_{T, n}$ is the $\log$-likelihood of the model, $\left\lvert S_{\hat{\theta}_n(\kappa)} \right\rvert$ is the size of this support, namely the number of active coefficients of $\hat{\theta}_n(\kappa)$. Compared to a classical \textsc{BIC} criterion (namely $\gamma=0$), an additional penalization is introduced. It that accounts for the number of possible active sets of the same size and is designed to better handle the high-dimensional setting. 

\paragraph{Optimization.} This paragraph is dedicated to the description of tools for solving the convex optimization problem given by Equation~\eqref{eq:optim}. The objective function is written as the sum of two functions. While the loss function $F_{T, n}(\cdot)$ is convex, differentiable and may even be smooth (in the case of least-squares contrast), the regularization function $\Omega(\cdot)$ may not be differentiable (the $\ell_1$-norm is not differentiable in 0). To this extent, to carry
out the minimization of such objective function, we consider first-order optimizations algorithm based
on proximal methods. We focus on two optimization algorithms: 
\begin{itemize}
    \item gradient descent (\textsc{GD}) (see \cite{Bach:2024}),
    \item accelerated gradient descent (\textsc{AGD}) (see \cite{Nesterov:1983}).
\end{itemize}
The \textsc{GD} algorithm is often regarded as the workhorse of first-order optimization, known for its simplicity and effectiveness across a broad range of applications. While straightforward in implementation, it can be slow to converge, especially in poorly conditioned settings. \textsc{AGD} refines this classic approach by incorporating a momentum term that strategically combines information from the two most recent iterates. Although this refinement introduces a slight additional computational cost, it is marginal, as it requires only a single gradient evaluation per iteration, as in standard \textsc{GD}. In exchange, it achieves a significantly faster convergence rate, often leading to a considerable reduction in the number of iterations required to achieve the desired level of accuracy.

In our case, both of these optimizers are coupled with the proximal operator to incorporate regularization. For the considered regularization, the corresponding proximal operators admit closed-form solutions.  In particular, in the case of Lasso regularization, the proximal operator reduces to the soft-thresholding function and theoretical properties of both optimizers, known in the literature as \textsc{ISTA} and \textsc{FISTA}, are well-studied and thoroughly documented (see \cite{Beck:2009}). 

Finally, a critical consideration in Gradient Descent is the selection of an appropriate step size. When the loss function has a Lipschitz continuous gradient, a commonly recommended step size is $1/L$, where 
$L$ denotes the Lipschitz constant of the gradient. In cases where this assumption does not hold, a backtracking line-search method can be employed to dynamically determine an effective step size at each iteration, ensuring stable and efficient convergence of the descent process (see \cite{Armijo:1966}).

\subsection{Multiclass classification}\label{subsec:classification}

Before introducing classification algorithms tailored for MHP path classification, let us first define the supervised multiclass classification model.

\paragraph{Multiclass classification model.} 
We consider the setting of $K$-multiclass classification with $K \geq 2$ the number of classes. Let $(\mathcal{T}_T, Y)$ be the couple of variables formed by the feature and its associated label. In this case, the feature $\mathcal{T}_T$ consists of time events of a MHP $N=(N_1, \dots, N_d)$ observed on $[0,T]$ whose dynamics of occurrence is governed by its label $Y \in [K]$. More precisely, conditional on $Y$, $N$ is a MHP of intensity $\lambda_{\theta_Y^*} = (\lambda_{\theta_Y^*,1}, \dots, \lambda_{\theta_Y^*,d})$. For each $j \in [d]$, the intensity function of the process $N_j$ depends on $Y$ and is given by:
$$\lambda_{\theta_Y^*,j}(t) = \mu_{Y,j}^* + \sum_{j'=1}^d \alpha_{Y,j,j'}^* \int_0^t \beta e^{-\beta(t-s)} \dd N_{j'}(s).$$

So in this model, the classes are discriminated according to the exogenous intensity vectors $(\mu_Y^*)_{Y \in [K]}$ and the interaction matrices $(\alpha_Y^*)_{Y \in [K]}$. This implies that classes are characterized by their underlying rate of spontaneous activity, as well as by the intrinsic structure of their interaction network. In the following, we assume that the parameter $\theta^* = (\mu_k^*, \alpha_k^*)_{k \in [K]}$ is unknown as well as the distribution of $Y$ which is denoted by $p^* = (p_k^*)_{k \in [K]}$. On the other hand, the decay rate $\beta$ is assumed to be known and identical for each class. Such a modeling assumption implies that the speed at which influences between network components vanish over time cannot be used as a benchmark to distinguish between classes. 

We follow the paradigm of supervised classification: although the law of the pair $(\mathcal{T}_T, Y)$ is unknown, we assume we have access to a learning sample $\mathcal{D}^{L}_n = \left\{\left(\mathcal{T}_T^{(1)}, Y^{(1)}\right), \dots, \left(\mathcal{T}_T^{(n)}, Y^{(n)}\right) \right\}$ of size $n$ which consists of independent copies of $(\mathcal{T}_T, Y)$. From training data the goal is to learn a decision rule $\hat{g}$, \textit{i.e.} a function of the data with values in [K], which accurately predicts the class to which an observation belongs.

\paragraph{\textsc{ERM} classifier.} As the distribution of the couple $(\mathcal{T}_T, Y)$ is unknown, it is not possible to calculate and therefore minimize the risk of misclassification. To mimic the Bayes classifier, which by definition minimizes this risk, a classic idea consists in minimizing the point error committed on each observation of the training sample over a chosen hypothesis space of classifier functions. This is the principle behind the empirical risk minimization (ERM) method, which has been covered abundantly in the literature (see \cite{Vapnik:1991}; \cite{Massart:2006}). 

Let us now present the \textsc{ERM} classification procedure tailored to classify Multivariate Hawkes Processes paths. This procedure was introduced in \cite{Dion-Blanc:2024}, see for more details on its principle and asymptotic properties. First, we estimate $p^*$ by its empirical counterpart $\hat{p}$. Then, rather than minimizing the empirical risk with the classic loss $0-1$, which leads to difficult optimization problems, we instead consider a convex surrogate (see \cite{Zhang:2004}; \cite{Bartlett:2006}). 

Let $\hat{\Theta}$ be the set of interest, $\theta \in \hat{\Theta}$, and $f_{\hat{p},\theta}$ its associated score function (see \cite{Dion-Blanc:2024} for more details). Consider the $L_2$-risk given by the following form:

\begin{equation}
\hat{\mathcal{R}}_2\left(f_{\hat{p}, \theta}\right) := \dfrac{1}{n} \sum_{i=1}^n \sum_{k = 1}^K \left(Z_k^{(i)}-f_{\hat{p}, \theta}^k\left(\mathcal{T}_T^{(i)}\right)\right)^2, \quad Z_k^{(i)} = 2\one_{\{Y_i=k\}}-1.
\end{equation}

Then, we define the estimator of $\theta^*$ as the minimizer of the empirical $L_2$-risk:
\begin{equation}\label{eq:optimERM}
\hat{\theta}^R \in \argmin_{\theta \in \hat{\Theta}} \hat{\mathcal{R}}_2(f_{\hat{p}, \theta}).
\end{equation}
From the estimator $\hat{\theta}^R$ of parameter $\theta^*$, we define the \textsc{ERM} classifier as follows
\begin{equation}\label{eq:classifERM}
\hat{g}(\mathcal{T}_T) \in  \argmax_{k \in \mathcal{Y}} f_{\hat{p}, \hat{\theta}^R}^k(\mathcal{T}_T)
\end{equation}

Having outlined the procedure, we now delve deeper into the practical aspects of solving the optimization problem presented in Equation~\eqref{eq:optimERM}.
To ensure the positivity of each coefficient, the minimization is conducted under inequality constraints, using a projected-type gradient descent algorithm. However, since this objective function is non-smooth and non-convex {\it w.r.t.} to the coefficients, its minimization demands careful consideration. 
In particular, tuning the step size in the descent process proves to be quite challenging. Furthermore, classical method such as backtracking line-search with Armijo-Wolfe condition cannot be used due to the piece-wise constant nature of the projection operator (see \cite{Ferry:2023}).

On the other hand, adaptive gradient methods, such as \textsc{AdaGrad} (see \cite{Duchi:2011}), have become widely adopted in large-scale optimization due to their ability to adjust the step size for each feature according to the geometry of the problem. In practice, \textsc{AdaGrad} is recognized as an efficient method in non-convex settings, particularly for training deep neural networks (see \cite{Gupta:2014}). In addition, some theoretical guarantees for the convergence of \textsc{AdaGrad} for non-convex functions have been established in the literature (see \cite{Ward:2020,Wang:2023}).

With this in mind, we use a parameter-free projected adaptive gradient descent method, in the likeness of \textsc{AdaGrad}, called \textsc{Free AdaGrad} and introduced in \cite{Chzhen:2023}. 
Compared with the classical algorithm, its key advantage lies in the fact that it is adaptive to the distance between the initialization and the optimum, and to the sum of the square norm of the gradients. The only input parameter of the
algorithm is $\gamma_0$, which can be interpreted as an initial lower-bound guess for the distance between the starting point and the optimum. In practice, we set $\gamma_0 = 0.1$. 

\paragraph{High-dimension.} Consider the high-dimensional framework, meaning the dimension of the network may be large \textit{w.r.t.} the number of observations. In such setting, the computational burden of the \textsc{ERM} procedure becomes substantial and the predictor $\hat{g}$ may not be consistent. Fortunately, high-dimensional data frequently exhibit a structure that is significantly more low-dimensional than their apparent complexity suggests. In this context, it is thus reasonable to postulate a sparse underlying structure and to endeavor to recover this lower-dimensional representation. Here, the sparsity hypothesis results in a sparse interaction matrix for each class. 

In line with this, \cite{Dion-Blanc:2024} introduced a \textsc{ERMLR} procedure whose methodology is the following. First, an initial interaction recovery step, by class, is performed using a Lasso-type estimator. Then, taking advantage of the estimation of the support of the adjacency matrix, by class, a refitting step based on a ERM criterion is carried out. In particular, the optimization of the $L_2$-risk is done on the following set of parameters:
$$\hat{\Theta}^R := \left\{\theta \in \hat{\Theta}, \ S_{\theta_k} = \hat{S}_k\right\}$$
where $\hat{S}_k$ is the estimated support of the $k$-th class given by the Lasso estimator. Hence, the minimization is performed on a set of parameters whose dimension is much smaller than $d^2$. The lasso step is significant as it selects effective interactions, thereby alleviating the complexity of the problem. On the other hand, an induced and undesired effect of $\ell_1$ penalization is the shrinkage of large coefficients. To bypass this issue, refitting strategies are commonly used and well covered in the literature, see for example \cite{Chzhen:2019}. Performing a refitting stage is, therefore, a natural progression, particularly in this context, as we employ a criterion specifically tailored for classification. The advantages of this approach are threefold: first, the selection of interactions enhances interpretability; second, it improves the classifier's performance; and third, it reduces the overall computational cost of the procedure. The \textsc{ERMLR} procedure is delineated comprehensively in Algorithm~\ref{alg:ermlr}.

\begin{algorithm}[H]
\SetAlgoLined

\caption{\textsc{ERMLR} procedure}\label{alg:ermlr}
\SetAlgoLined
\SetKwInput{Input}{Input}
\SetKwInput{Init}{Initialization}
\SetKwInput{Output}{Output}

\Input{$\mathcal{D}_n$} 

\MyComment{Class weight estimation}
\For{$k=1, \dots, K$}{
    $\hat{p}_k = \frac{1}{n}\sum_{i=1}^n \one_{\{Y^i=k\}}$\;
}
\MyComment{Support recovery}
\For{$k=1, \dots, K$}{
    Choose $\hat{\kappa}_k \in \argmin_{\kappa \in \Delta} \text{EBIC}_1(\kappa)$ \;
    Solve $\hat{\theta}_k \in \argmin_{\theta \in \mathbb{R}^{d \times d+1}} \left\{R_{T, n}(\theta) + \hat{\kappa}_k \norm{\theta}_1\right\}$ using \textsc{FISTA}\;
    Get $\hat{S}_k = \left\{\hat{\theta}_{k,j,j'} \neq 0, \ (j, j') \in [d]\right\}$
}
\MyComment{ERM reffiting}
Solve $\hat{\theta}^R \in \argmin_{\theta \in \hat{\Theta}^R} \hat{\mathcal{R}}_2(f_{\hat{p}, \theta})$ using \textsc{Free AdaGrad} starting from $(\hat{\theta}_k)_{k \in [K]}$ \;
Get $\hat{g}(\cdot) \in  \argmax_{k \in \mathcal{Y}} f_{ \hat{\theta}^R}^k(\cdot)$ \;
\Output{$\hat{g}(\cdot)$}
\end{algorithm}

\section{Package presentation and functionality}\label{sec:implementation}

This section begins with an overview of the global architecture and features of the package in Section~\ref{subsec:code_overview}. This is followed by a detailed explanation of the implementation of the statistical methods in Section~\ref{subsec:code_simu}, \ref{subsec:code_infer}, \ref{subsec:code_classif}, closely mirroring the structure of Section~\ref{sec:theory_hawkes}. In these sections, comprehensive and intuitive guides are provided on how to apply these methods in practice, complemented by insightful examples for better illustration.

\subsection{Package overview}
\label{subsec:code_overview}

\begin{figure}
    \centering
    \begin{forest}
      pic dir tree,
      where level=0{}{
        directory,
      },
      [Sparklen
        [lib, label=right:{\small Contains \proglang{C++} and \pkg{SWIG} interface files}
            [include
            ]
            [interface
            ]
            [src
            ]
        ]
        [Sparklen, label=right:{\small \proglang{Python} package folder}
            [...
            ]
        ]
        [setup.py, file, label=right:{\small Build and installation script}
        ]
        [README.md, file, label=right:{\small Documentation}
        ]
      ]
    \end{forest}
    \caption{Tree structure of \pkg{Sparklen} library directory.}
    \label{fig:dir_tree}
\end{figure}
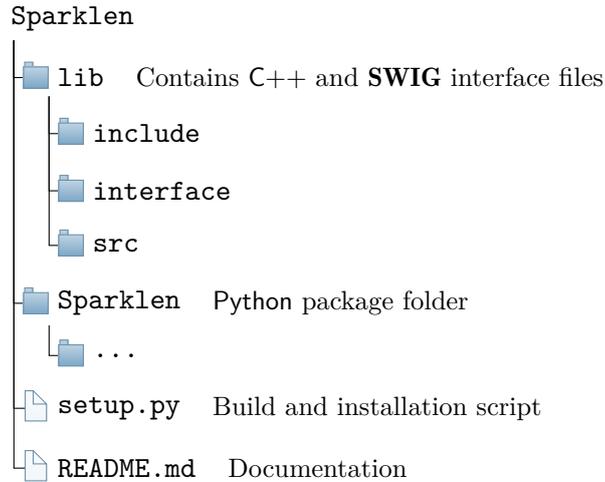

\paragraph{Global architecture.} The directory structure of the package is thoughtfully organized to clearly compartmentalize its various components. At the core, the \texttt{lib/} folder houses \proglang{C++} header and source files, along with \pkg{SWIG} interface files. These elements are not intended for direct user interaction; rather, they serve the primary purpose of enabling rapid computation, allowing the package to perform complex calculations efficiently while abstracting away the underlying \proglang{C++} implementation. All user-ended \proglang{Python} code resides exclusively within the \texttt{sparklen/} directory, which encompasses the suite of modules and sub-modules offered. This directory serves as the primary interface for users, encapsulating the entire functionality of the package. It is meticulously organized to facilitate ease of use and maintainability, allowing users to access and possibly extend the package's capabilities if needed. The \texttt{setup.py} script plays a pivotal role in the package's architecture by orchestrating the build process. It ensures that \pkg{SWIG} compiles the interface files correctly into \proglang{Python} extension modules, making the high-performance \proglang{C++}  functionalities accessible from \proglang{Python}. This script also manages dependencies and installation requirements, streamlining the installation process for users. Additionally, a \texttt{README.md} file is included to offer documentation, detailing installation instructions, link to usage examples, and other resources to assist users in effectively navigating the package. A visual representation of this structure is given in~Figure \ref{fig:dir_tree}.

\paragraph{Features of the \pkg{Sparklen} package.} We will now delve into the features of the package, encapsulated within the \texttt{sparklen/} directory as a collection of modules and sub-modules. Each module is named after the package, with sub-modules separated from the parent module name by a dot, following Python's standard attribute syntax. This naming convention not only promotes clarity, but also makes it easier to navigate and use. There are essentially four modules: \texttt{sparklen.hawkes}, \texttt{sparklen.optim}, \texttt{sparklen.prox}, \texttt{sparklen.calibration}. 

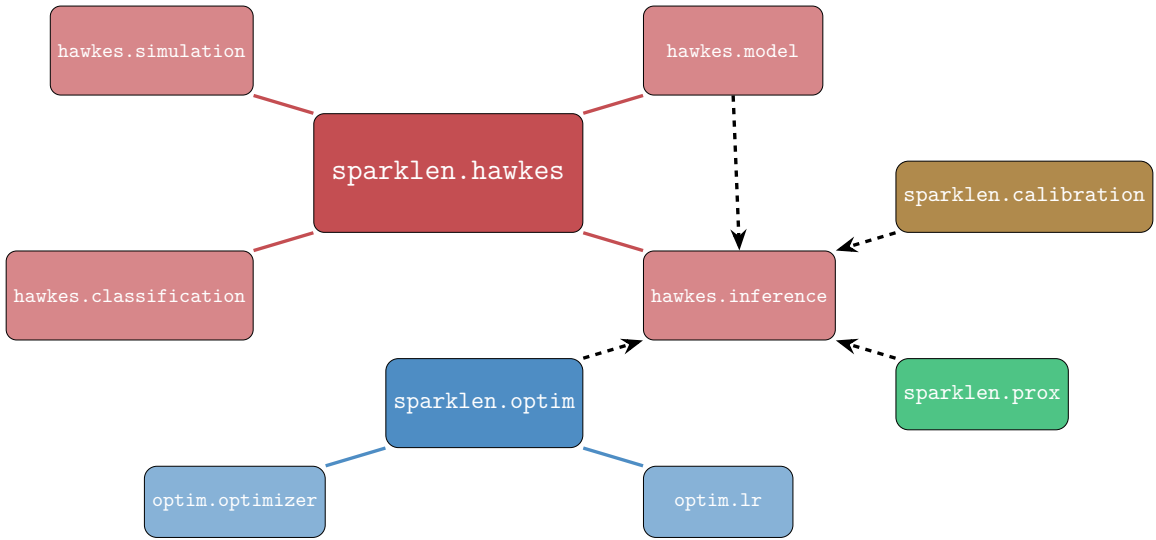
\begin{figure}[ht]
    \centering
    \resizebox{0.99\textwidth}{!}{

\begin{tikzpicture}[
    hawkesmod/.style={draw, fill=red, rounded corners=5pt, minimum width=4.5cm, minimum height=2cm, text centered, text=white, font=\Large}, 
    hawkessubmod/.style={draw, fill=lightred, rounded corners=5pt, minimum width=3cm, minimum height=1.5cm, text centered, text=white, font=\small},
    optimmod/.style={draw, fill=blue, rounded corners=6pt, minimum width=3.0cm, minimum height=1.5cm, text centered, text=white, font=\large}, 
    optimsubmod/.style={draw, fill=lightblue, rounded corners=6pt, minimum width=2.5cm, minimum height=1.2cm, text centered, text=white,font=\small},
    proxmod/.style={draw, fill=green, rounded corners=6pt, minimum width=2.5cm, minimum height=1.2cm, text centered, text=white}, 
    penmod/.style={draw, fill=brown, rounded corners=6pt, minimum width=2.5cm, minimum height=1.2cm, text centered, text=white}, 
    hierlinehawkes/.style={-, ultra thick, color=red}, 
    hierlineoptim/.style={-, ultra thick, color=blue},
    deparrow/.style={->, dashed, ultra thick, color=black, >=Stealth}, 
    node distance=0.3cm and 1cm 
]

\node[hawkesmod] (mod1) {\texttt{sparklen.hawkes}};

\node[hawkessubmod] (sub1) [above left=of mod1] {\texttt{hawkes.simulation}};
\node[hawkessubmod] (sub2) [above right=of mod1] {\texttt{hawkes.model}};
\node[hawkessubmod] (sub3) [below right=of mod1] {\texttt{hawkes.inference}};
\node[hawkessubmod] (sub4) [below left=of mod1] {\texttt{hawkes.classification}};

\node[optimmod] (mod2) [below left=of sub3] {\texttt{sparklen.optim}};
\node[penmod] (mod3) [above right=of sub3] {\texttt{sparklen.calibration}};
\node[proxmod] (mod4) [below right=of sub3] {\texttt{sparklen.prox}};

\node[optimsubmod] (sub2_1) [below left=of mod2] {\texttt{optim.optimizer}};
\node[optimsubmod] (sub2_2) [below right=of mod2] {\texttt{optim.lr}};

\draw[hierlinehawkes] (mod1.north west) -- (sub1.south east);
\draw[hierlinehawkes] (mod1.north east) -- (sub2.south west);
\draw[hierlinehawkes] (mod1.south east) -- (sub3.north west);
\draw[hierlinehawkes] (mod1.south west) -- (sub4.north east);

\draw[hierlineoptim] (mod2.south west) -- (sub2_1.north east);
\draw[hierlineoptim] (mod2.south east) -- (sub2_2.north west);

\draw[deparrow] (mod2.north east) -- (sub3.south west); 
\draw[deparrow] (mod3.south west) -- (sub3.north east); 
\draw[deparrow] (mod4.north west) -- (sub3.south east); 
\draw[deparrow] (sub2.south) -- (sub3.north); 

\end{tikzpicture}
    }
    \caption{The modules and sub-modules of \pkg{Sparklen} package are shown, in the form of a flowchart, along with potential dependencies specified by dashed arrows.}
    \label{fig:flowchart}
\end{figure}

The flagship and most crucial module is \texttt{sparklen.hawkes}. Such a module integrates all functionalities associated with Hawkes processes, each organized within the following four sub-modules. The sub-module \texttt{sparklen.hawkes.simulation} is responsible for simulating Hawkes processes. The \texttt{sparklen.hawkes.model} manages the calculations related to the model, such as the evaluation of the loss or the calculation of the gradient for the log-likelihood and least-squares models. The \texttt{sparklen.hawkes.inference} contains all the tools for carrying out Hawkes processes inference. In particular, it is based on the previous sub-module and uses the \texttt{sparklen.optim} modules for the optimization phase, \texttt{sparklen.prox} to manage the integration of a regularization term and \texttt{sparklen.calibration} to choose the penalty constant. Finally in \texttt{sparklen.hawkes.classification} are implemented tools for classification. See the following sections for more details on these sub-modules. A visual representation of modules, sub-modules and potential dependencies is given in Figure~\ref{fig:flowchart}.

\paragraph{\proglang{C++}/\proglang{Python} binding.} An appealing feature of this package lies in its ability to leverage the strengths of both \proglang{C++} and \proglang{Python}. By implementing the computationally intensive components in \proglang{C++} (like gradient evaluation), we ensure that the most resource-heavy tasks are executed with optimal efficiency, taking full advantage of \proglang{C++}'s speed and performance. At the same time, the \proglang{C++} code is seamlessly wrapped in \proglang{Python} using \pkg{SWIG}, which provides a user-friendly, Pythonic interface. This approach offers the best of both worlds: users benefit from the fast, low-level computations of \proglang{C++}, while interacting with high-level, intuitive, flexible \proglang{Python} objects that are easy to manipulate and integrate into their workflows. This dual-language strategy makes it possible to deal with resource-intensive applications (\textit{e.g.} inference of high-dimensional Hawkes processes) while guaranteeing ease of use. A concise yet comprehensive explanation of the mechanisms underpinning the interface between \proglang{C++} and \proglang{Python} is provided in~Appendix \ref{app:mecanism}. 

In the following, we present the main features of the module \texttt{sparklen.hawkes}, namely  simulation, inference and classification. This part complements Section~\ref{sec:theory_hawkes} by shedding light on how these tools are implemented in practice, with examples provided for added clarity. In each example the workflow, illustrated with code snippets, is interspersed with explanatory comments to enhance explicitness.

\subsection{Implementation of the simulation framework}
\label{subsec:code_simu}

\paragraph{Description.} Everything related to the simulation of Hawkes processes is contained in the sub-module \texttt{sparklen.hawkes.simulation}. The centerpiece of this sub-module is the \texttt{SimuHawkesExp} class. Each object in this class is instantiated by giving the following parameters:
\begin{itemize}
    \item \texttt{mu}: exogenous intensities (\texttt{ndarray} of shape \((d,)\)), 
    \item \texttt{alpha}: interaction matrix (\texttt{ndarray} of shape \((d, d)\)), 
    \item \texttt{beta}: common decay scalar (\texttt{float}),
    \item \texttt{end\_time}: horizon time (\texttt{float}),
    \item \texttt{n\_samples}: number of repeated paths (\texttt{int}),
    \item \texttt{random\_state}: seed for random number generation (\texttt{int} or \texttt{None}).
\end{itemize}

Once this has been done, several methods of interest can be called. The main method is \texttt{simulate()}, which simulates the events of the considered Hawkes process. A brief list of the usable methods of this class is given in the following Table~\ref{tab:method_simu}.

\begin{table}[ht]
\centering
\begin{tabular}{ll}
\toprule
\textbf{Method} & \textbf{Description} \\ \midrule
\texttt{simulate()} & \textit{Simulate} process events. \\
\texttt{compensator(t)} & \textit{Compute} the compensator evaluated in \texttt{t}. \\
\texttt{spectral\_radius()} & \textit{Return} the spectral radius of the interaction matrix. \\
\texttt{print\_info()} & \textit{Display} key details about the object. \\
\bottomrule
\end{tabular}
\caption{Methods of the \texttt{SimuHawkesExp} class from the \pkg{Sparklen} package.}
\label{tab:method_simu}
\end{table}

\paragraph{Example 1: simulation of a MHP.}\label{ex:simulation}

\begin{figure}[htbp]
    \centering
    \includegraphics[width=0.6\textwidth]{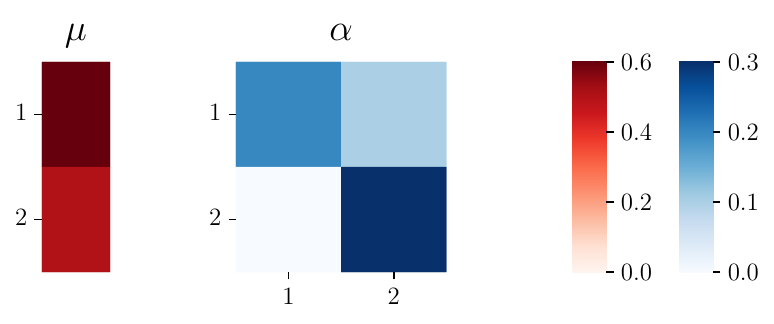}
    \caption{Heatmap visualization of $\theta^* = (\mu^*, \alpha^*)$, where $\mu^*$ values are shown in red (gradient on the left colorbar) and $\alpha^*$ values are shown in blue (gradient on the right colorbar).}
    \label{fig:E1_true_values}
\end{figure}

The following provides a comprehensive guide to using the \texttt{SimuHawkesExp} class to simulate the events of a MHP. In this example, we consider a $2$-MHP whose parameters are given in Figure~\ref{fig:E1_true_values}.

It all starts with importation of the class \texttt{SimuHawkesExp} from the \texttt{simulation} sub-module. 

\begin{CodeChunk}
\begin{CodeInput}
>>> from sparklen.hawkes.simulation import SimuHawkesExp
\end{CodeInput}
\end{CodeChunk}

The next step is to instantiate an object of this class. At this stage, the parameters of the specified MHP, along with the desired time horizon and number of repetitions, must be provided. In this example, we set $\beta=3.0$, $T = 5.0$ and, for simplicity, choose $n = 3$ repetitions. In addition, note that the seed is fixed, ensuring reproducible output across multiple runs of this example code.

\begin{CodeChunk}
\begin{CodeInput}
>>> hawkes = SimuHawkesExp(
...     mu=mu, alpha=alpha, beta=beta,
...     end_time=T, n_samples=n,
...     random_state=8)
\end{CodeInput}
\end{CodeChunk}

With the object instantiated, we can invoke the \texttt{simulate} method to generate realizations. 

\begin{CodeChunk}
\begin{CodeInput}
>>> hawkes.simulate()
\end{CodeInput}
\end{CodeChunk}

Having done this, we can access the generated data via the \texttt{timestamps} property.

\begin{CodeChunk}
\begin{CodeInput}
>>> hawkes.timestamps
\end{CodeInput}
\begin{CodeOutput}
[[array([0.53476798, 1.20676073, 1.86374452, 2.18940937, 2.39482727]),
  array([0.96932274, 1.155102  ])],
 [array([2.26133264, 3.08363972]),
  array([0.48054217, 0.70050504, 1.64644688, 2.26772207, 2.48926523])],
 [array([0.67427111, 2.29518797, 3.05520982]), 
  array([1.86160981])]]
\end{CodeOutput}
\end{CodeChunk}

\begin{figure}[htbp]
    \centering
    \includegraphics[width=0.6\textwidth]{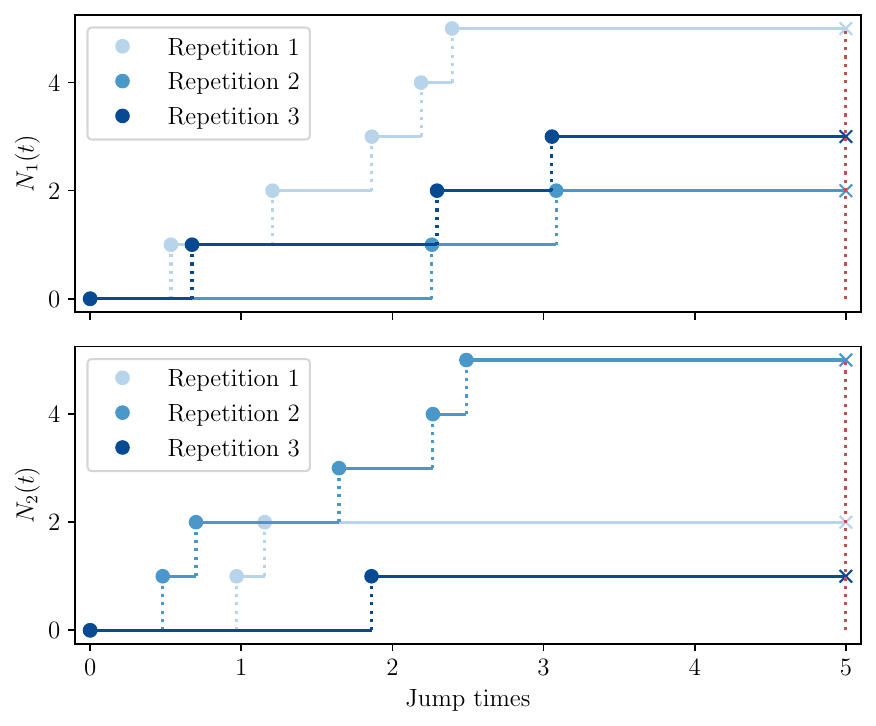}
    \caption{Visualization of the simulated timestamps of the 2-MHP. On top, the repeated paths of the first component of the network; on the bottom, the repeated paths of the second component. The repetition of paths within each dimension is color-coded using a gradient of blue.}
    \label{fig:E1_repeated_paths}
\end{figure}

An illustration of these repeated paths is supplied in Figure~\ref{fig:E1_repeated_paths}.

\subsection{Implementation of the inference framework}
\label{subsec:code_infer}

\paragraph{Description.} The \pkg{Sparklen}'s inference tools are housed within the dedicated sub-module \texttt{sparklen.hawkes.inference}. 
This module relies on the \texttt{sparklen.optim}, \texttt{sparklen.prox}, and \texttt{sparklen.calibration} modules (see Figure~\ref{fig:flowchart}) to effectively carry out the minimization of the problem given by Equation~\eqref{eq:optim}. The flagship \texttt{LearnerHawkesExp} class operates as a central unit, providing a unified interface that orchestrates the interplay between these three modules. This implementation directly mirrors the modular structure presented in Section~\ref{subsec:inference}, granting users access to the same extensive range of options. By abstracting the complexity of managing these interconnected modules, it ensures a smooth and seamless workflow. At the same time, the design maintains flexibility, allowing users to customize the estimation procedure through intuitive string-based arguments and configuration dictionaries. 
To sum up, the benefits of this approach are two-fold: it simplifies the user experience while ensuring the module remains adaptable to a wide variety of application needs. The following options are available:

\begin{itemize}
    \item \texttt{loss}: 'least-squares' or 'log-likelihood';
    \item \texttt{penalty}: 'none', 'lasso', 'ridge' or 'elasticnet';
    \item \texttt{kappa\_choice}: 'cv', 'bic' or 'ebic';
    \item \texttt{optimizer}: 'gd' or 'agd';
    \item \texttt{lr\_scheduler}: 'lipschitz' or 'backtracking'.
\end{itemize}

Each object of this class is instantiated by specifying the configuration choices, along with additional arguments where applicable (e.g., the number of folds for cross-validation if 'cv' is chosen, or the hyperparameter $\gamma$ if the 'ebic' criterion is selected). After instantiation, the user must call the \texttt{fit()} method to fit the model to the data. Once the model has been fitted, several methods of interest can be invoked. A brief overview of the available methods for this class is provided in Table~\ref{tab:method_estim}.

\begin{table}[ht]
\centering
\begin{tabular}{ll}
\toprule
\textbf{Method} & \textbf{Description} \\ \midrule
\texttt{fit(data, end\_time)} & \textit{Fit} Hawkes model to data. \\
\texttt{score(data, end\_time)} & \textit{Compute} score metric. \\
\texttt{plot\_estimated\_values()} & \textit{Plot} estimated parameter values. \\
\texttt{plot\_estimated\_support()} & \textit{Plot} estimated parameter support. \\
\texttt{print\_info()} & \textit{Display} key details about the object. \\
\bottomrule
\end{tabular}
\caption{Methods of the \texttt{LearnerHawkesExp} class from the \pkg{Sparklen} package}
\label{tab:method_estim}
\end{table}

\begin{figure}[htbp]
    \centering
    \includegraphics[width=0.7\textwidth]{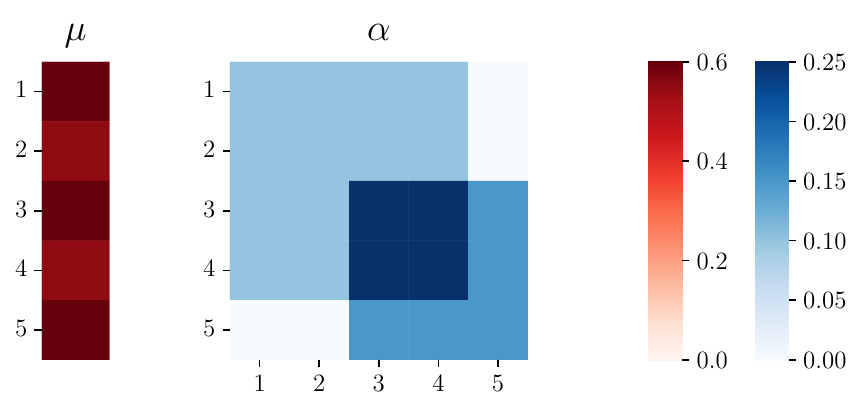}
    \caption{Heatmap visualization of $\theta^* = (\mu^*, \alpha^*)$, where $\mu^*$ values are shown in red (gradient on the left colorbar) and $\alpha^*$ values are shown in blue (gradient on the right colorbar).}
    \label{fig:E2_true_values}
\end{figure}

\paragraph{Example 2: inference of a MHP.} The following provides a comprehensive guide to using the \texttt{LearnerHawkesExp} class to get an estimation of the true parameter $\theta^*$ of a MHP. In this example, we consider a $5$-MHP whose parameter is given in Figure~\ref{fig:E2_true_values}.

Before getting started, we need to import the \texttt{SimuHawkesExp} responsible for the simulation of synthetic data and \texttt{LearnerHawkesExp} to carry out the inference task. 

\begin{CodeChunk}
\begin{CodeInput}
>>> from sparklen.hawkes.simulation import SimuHawkesExp
>>> from sparklen.hawkes.inference import LearnerHawkesExp
\end{CodeInput}
\end{CodeChunk}

The initial step involves simulating the repeated paths of the corresponding MHP, which will serve as the training set for subsequent inference. For this demonstration, we choose $\beta = 3.0$, we use a training sample of size $n = 1000$ and a time horizon of $T = 5.0$. Additional details on this step can be found in Example~\ref{ex:simulation}.

\begin{CodeChunk}
\begin{CodeInput}
>>> hawkes = SimuHawkesExp(
...     mu=mu, alpha=alpha, beta=beta,
...     end_time=T, n_samples=n,
...     random_state=4)

>>> hawkes.simulate()

>>> data = hawkes.timestamps
\end{CodeInput}
\end{CodeChunk}

The next step is to instantiate an object of this class. At this stage, we must give the decay parameter, and the chosen options of the learner. For readability, the \texttt{verbose\_bar} argument is set to 'False'. If set to 'True', a progress bar will be displayed to monitor the optimization progress. The \texttt{verbose} argument, on the other hand, controls whether recorded information during the optimization phase is printed in the end. If set to 'True', details will be displayed every \texttt{print\_every} iterations.

\begin{CodeChunk}
\begin{CodeInput}
>>> learner = LearnerHawkesExp(
...     decay=beta, loss="least-squares", penalty="none", 
...     optimizer="agd", lr_scheduler="backtracking", 
...     max_iter=200, tol=1e-5, verbose_bar=False,
...     verbose=True, print_every=10, record_every=10)
\end{CodeInput}
\end{CodeChunk}

With the object instantiated, we can fit the model using the given training data. The output includes information recorded during optimization, the termination status, and the total duration of the procedure. 

\begin{CodeChunk}
\begin{CodeInput}
>>> learner.fit(data, T)
\end{CodeInput}

\begin{CodeOutput}
Optimization completed. Convergence achieved after 24 iterations.

Time elapsed: 0.40 seconds.
+-------------+----------+-------------+
|   Iteration |     Loss |   Tolerance |
+=============+==========+=============+
|           0 | -12.7803 | 0.0863339   |
+-------------+----------+-------------+
|          10 | -13.5298 | 0.000408739 |
+-------------+----------+-------------+
|          20 | -13.5624 | 2.35301e-05 |
+-------------+----------+-------------+
\end{CodeOutput}
\end{CodeChunk}

Having done this, we can obtain the estimated parameters by calling the \texttt{estimated\_params} property method.

\begin{CodeChunk}
\begin{CodeInput}
>>> learner.estimated_params
\end{CodeInput}

\begin{CodeOutput}
array([
    [0.56945931, 0.12918584, 0.09980257, 0.1054643 , 0.10675509, 0.00803352],
    [0.55539019, 0.11320206, 0.09733585, 0.08572127, 0.10494816, 0.00640289],
    [0.63591031, 0.10224589, 0.08963574, 0.24939751, 0.24530404, 0.15960453],
    [0.56215877, 0.0958248 , 0.08496727, 0.24182603, 0.24653997, 0.15607922],
    [0.59766823, 0.        , 0.        , 0.13851235, 0.13871558, 0.1488716 ]])
\end{CodeOutput}
\end{CodeChunk}

\begin{figure}[htbp]
    \centering
    \includegraphics[width=0.7\textwidth]{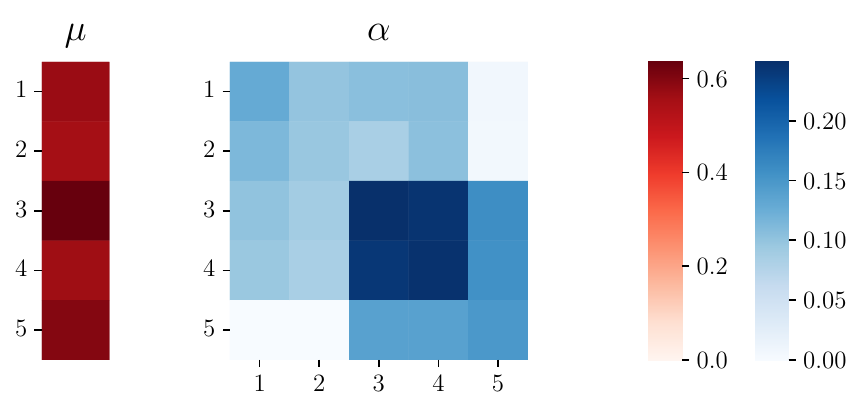}
    \caption{Heatmap visualization of $\hat{\theta} = (\hat{\mu}, \hat{\alpha})$, where $\mu^*$ values are shown in red (gradient on the left colorbar) and $\alpha^*$ values are shown in blue (gradient on the right colorbar).}
    \label{fig:E2_estimated_values}
\end{figure}

Additionally, for better visualization, we can plot the estimated values of the parameter by calling the method \texttt{plot\_estimated\_values()}, represented in Figure~\ref{fig:E2_estimated_values}.

\begin{CodeChunk}
\begin{CodeInput}
>>> learner.plot_estimated_values()
\end{CodeInput}
\end{CodeChunk}

As shown in Figure~\ref{fig:E2_estimated_values}, the estimation procedure provides a quality estimate of the true parameter $\theta^*$. The visualization demonstrates that the estimated parameter aligns well with the ground truth.

One can also get the score on training data by calling the \texttt{score()} method:

\begin{CodeChunk}
\begin{CodeInput}
>>> learner.score(data, T)
\end{CodeInput}

\begin{CodeOutput}
-13.56
\end{CodeOutput}
\end{CodeChunk}

\begin{figure}[htbp]
    \centering
    \includegraphics[width=0.7\textwidth]{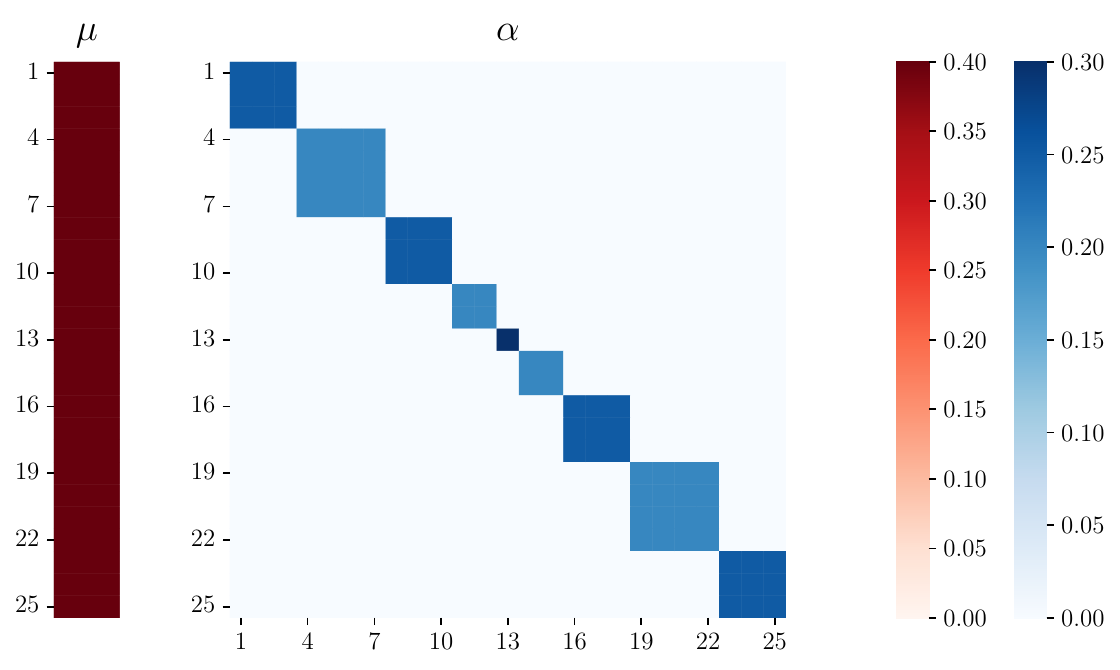}
    \caption{Heatmap visualization of $\theta^* = (\mu^*, \alpha^*)$, where $\mu^*$ values are shown in red (gradient on the left colorbar) and $\alpha^*$ values are shown in blue (gradient on the right colorbar). The sparsity rate of the matrix $\alpha^*$, \textit{i.e.} the proportion of zero coefficients, is equal to 0.88.}
    \label{fig:E3_true_values}
\end{figure}

\paragraph{Example 3: inference of a sparse high-dimensional MHP.} Now, we shift our focus to a high-dimensional MHP where the interaction matrix exhibits a sparse structure. This example provides a detailed walkthrough for using the LearnerHawkesExp class, showcasing the specialized options designed for such cases. We considrer a 25-dimensional MHP, with the parameter configuration shown in Figure~\ref{fig:E3_true_values}. 

First, as before, we need to import the classes and simulate the training dataset. For details on these steps, refer to the two examples above. We choose $\beta=3.0$, $n=250$ and $T=5.0$. In this example, the high-dimensional ratio, given by $d/n$, is equal to 0.1

Next, we need to create an instance of the class \texttt{LearnerHawkesExp} and specify the options for this learner. In this context, we opt for Lasso regularization, and we compare the tuning of the penalization constant being done using \textsc{EBIC} with $\gamma=1.0$ and using \textsc{CV} with a number of folds equal to $10$.

\begin{CodeChunk}
\begin{CodeInput}
>>> learner_lasso_cv = LearnerHawkesExp(
...     decay=beta, loss="least-squares", 
...     penalty="lasso", kappa_choice="cv"
...     optimizer="agd", lr_scheduler="backtracking", 
...     max_iter=200, tol=1e-5, 
...     verbose_bar=False, verbose=False,
...     cv=10)

>>> learner_lasso_ebic = LearnerHawkesExp(
...     decay=beta, loss="least-squares", 
...     penalty="lasso", kappa_choice="ebic"
...     optimizer="agd", lr_scheduler="backtracking", 
...     max_iter=200, tol=1e-5, 
...     verbose_bar=False, verbose=False,
...     gamma=1.0)
\end{CodeInput}
\end{CodeChunk}

Then, we can fit the model using the provided training data. 

\begin{CodeChunk}
\begin{CodeInput}
>>> learner_lasso_cv.fit(data, T)

>>> learner_lasso_ebic.fit(data, T)
\end{CodeInput}
\end{CodeChunk}

One can also get the score on training data by calling the \texttt{score()} method:

\begin{CodeChunk}
\begin{CodeInput}
>>> learner_lasso_cv.score(data, T)

>>> learner_lasso_ebic.score(data, T)
\end{CodeInput}

\begin{CodeOutput}
-66.83

-66.29
\end{CodeOutput}
\end{CodeChunk}

\begin{figure}[ht]
    \centering
    \begin{subfigure}[b]{0.49\textwidth}
        \centering
        \includegraphics[width=\linewidth]{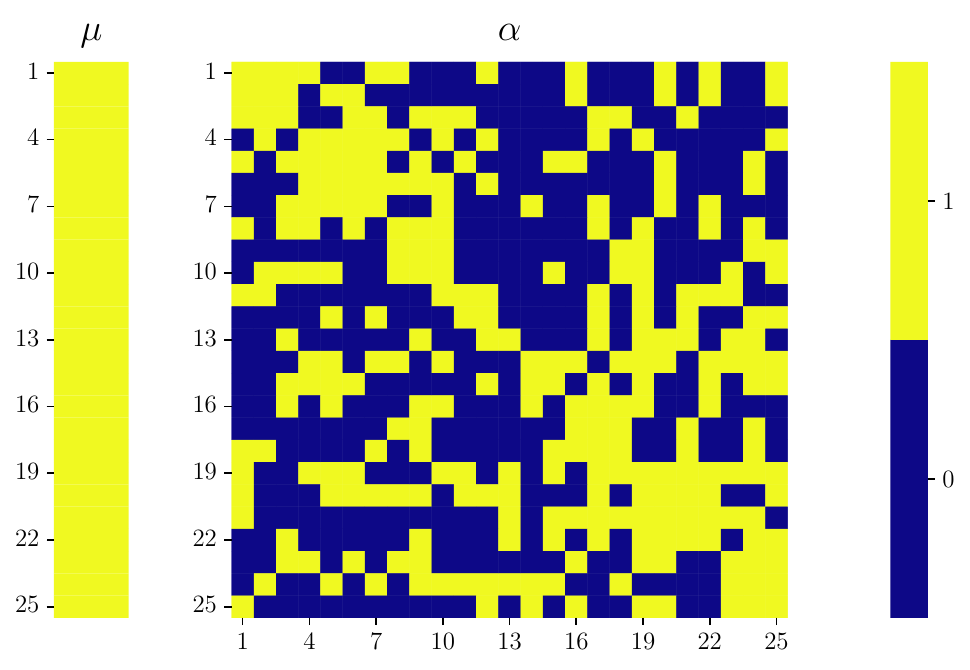}
        \caption{CV}
    \end{subfigure}
    \hfill
    \begin{subfigure}[b]{0.49\textwidth}
        \centering
        \includegraphics[width=\linewidth]{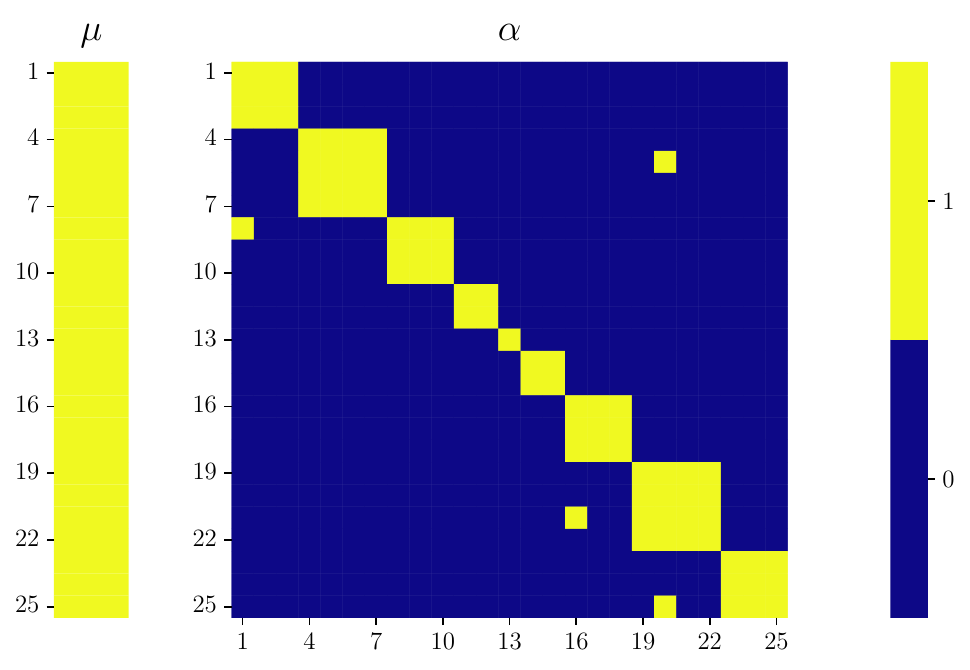}
        \caption{EBIC}
    \end{subfigure}
    \caption{Estimated support with Lasso regularization.}
    \label{fig:E3_lasso_estimated_support}
\end{figure}

In the context of sparse MHP in a high-dimensional setting, a key consideration is the ability to recover the true support of $\theta^*$. To demonstrate the efficacy of the Lasso estimator in interaction selection, we can visualize the estimated support by invoking the method \texttt{plot\_estimated\_support()}, represented in Figure~\ref{fig:E3_lasso_estimated_support}.

\begin{CodeChunk}
\begin{CodeInput}
>>> learner_lasso_cv.plot_estimated_support()

>>> learner_lasso_ebic.plot_estimated_support()
\end{CodeInput}
\end{CodeChunk}

As it can be seen in Figure~\ref{fig:E3_lasso_estimated_support}, the Lasso estimator, when using \textsc{CV} to select the penalization constant, tends to be overly conservative and insufficiently penalizes, failing to identify the true support. In contrast, tuning 
$\kappa$ using \textsc{EBIC} enables the Lasso estimator to successfully recover the underlying block structure of the interaction matrix.

\subsection{Implementation of the classification framework}
\label{subsec:code_classif}

\paragraph{Description.} The \pkg{Sparklen}'s classification tools are housed within the dedicated sub-module \texttt{n.hawkes.classification}. The first element is the \texttt{make\_classification} function responsible for the generation of a random $K$-class classification problem. The arguments of this function are the following:
\begin{itemize}
    \item \texttt{bold\_mu}: exogenous intensities of each class (\texttt{ndarray} of shape \((K, d,)\)),
    \item \texttt{bold\_alpha}: interaction matrix of each class (\texttt{ndarray} of shape \((K, d, d)\)),
    \item \texttt{beta}: common decay scalar (\texttt{float}),
    \item \texttt{end\_time}: horizon time (\texttt{float}),
    \item \texttt{n\_samples}: the number of samples (\texttt{int}),
    \item \texttt{n\_classes}: the number of classes (\texttt{int}),
    \item \texttt{weights}:  proportions of samples assigned to each class (\texttt{ndarray} of shape \((K, )\)),
    \item \texttt{random\_state}: seed for random number generation (\texttt{int} or \texttt{None}).
\end{itemize}

The rest of the module consists of tools for performing classification. In particular, the two classifier classes are the \texttt{ERMCLassifier} and the \texttt{ERMLRCLassifier} that implement the \textsc{ERM} and \textsc{ERMLR} classifier presented in Section~\ref{subsec:classification}. The core implementation logic is shared between the two classes and both are designed to follow the \pkg{scikit-learn} API (see \cite{Pedregosa:2011} and \cite{Buitinck:2013}). A list of usable methods of these class is given in Table~\ref{tab:method_classif}. 

\begin{table}[ht]
\centering
\begin{tabular}{ll}
\toprule
\textbf{Method} & \textbf{Description} \\ \midrule
\texttt{fit(X, y, end\_time)} & \textit{Build} a classifier from the training set (X, y). \\
\texttt{predict(X, end\_time)} & \textit{Predict} class for X. \\
\texttt{predict\_proba(X, end\_time)} & \textit{Predict} class probabilities for X. \\
\texttt{score(X, y, end\_time)} & \textit{Return} the mean accuracy on the given test set (X, y). \\
\texttt{print\_info()} & \textit{Display} key details about the object. \\
\bottomrule
\end{tabular}
\caption{Usable methods shared by the \texttt{ERMCLassifier} and \texttt{ERMLRCLassifier} class from the \pkg{Sparklen} package.}
\label{tab:method_classif}
\end{table}

\begin{figure}[htbp]
    \centering
    \includegraphics[width=0.7\textwidth]{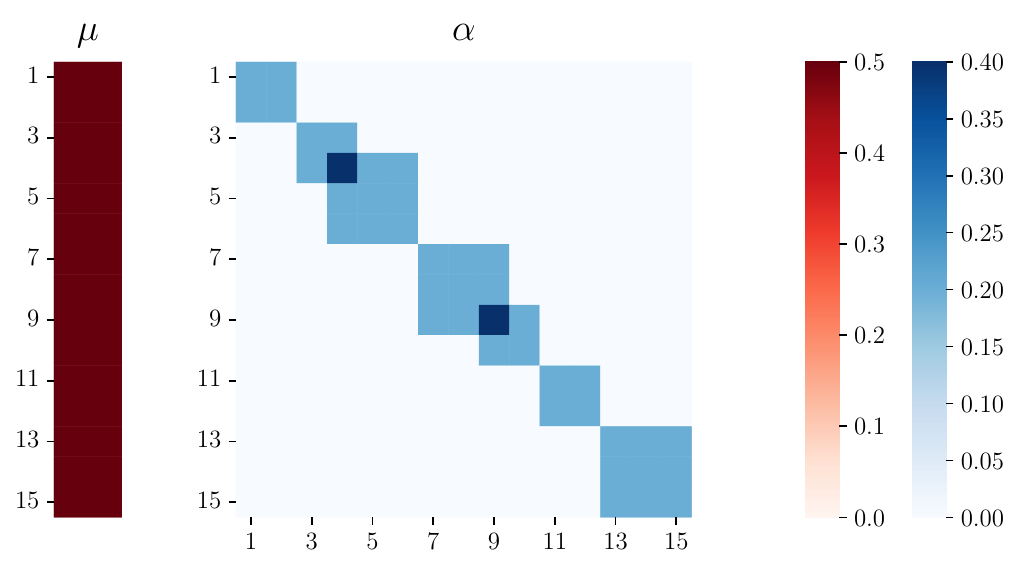}
    \caption{Heatmap visualization of $\theta^* = (\mu^*, \alpha^*)$, where $\mu^*$ values are shown in red (gradient on the left colorbar) and $\alpha^*$ values are shown in blue (gradient on the right colorbar).}
    \label{fig:E4_true_values}
\end{figure}

\paragraph{Example 4: classification of a MHP.} To illustrate the performance of both classifiers, we consider the following classification model with $K=3$ of a $15$-MHP. We represent the parameter associated to $k=2$ in Figure~\ref{fig:E4_true_values}. The parameters of the two other classes are obtained by interchanging the blocks of the interaction matrix. All three classes have the same exogenous intensity vector. Finally, we choose $\beta=3.0$, $n=600$, $T=5.0$ and the seed is fixed. 

To begin with, the function and class need to be imported. 

\begin{CodeChunk}
\begin{CodeInput}
>>> from sparklen.hawkes.classification import make_classification
>>> from sparklen.hawkes.classification import ERMCLassifier, ERMLRCLassifier
\end{CodeInput}
\end{CodeChunk}

The first step consists of generating the training and testing datasets. This can be achieved using the \texttt{make\_classification()} method, followed by splitting the data evenly, with 50\% allocated for training and the remaining 50\% for testing.

\begin{CodeChunk}
\begin{CodeInput}
>>> X, y = make_classification(
...     bold_mu=bold_mu, bold_alpha=bold_alpha, 
...     beta=beta, end_time=T,
...     n_samples=n, n_classes=K
...     random_state=4)

>>> X_train, X_test, y_train, y_test = train_test_split(
...     X, y, test_size=0.5, random_state=4)
\end{CodeInput}
\end{CodeChunk}

Then, we need to instantiate the classifier-type objects and to specify the chosen arguments.

\begin{CodeChunk}
\begin{CodeInput}
>>> clf_erm = ERMCLassifier(
...     decay=beta, gamma0=0.1, 
...     max_iter=500, tol=1e-6)

>>> clf_ermlr = ERMLRCLassifier(
...     decay=beta, gamma0=0.1, 
...     max_iter=500, tol=1e-6)
\end{CodeInput}
\end{CodeChunk}

Finally we can fit the model according to the given training set (X, y).

\begin{CodeChunk}
\begin{CodeInput}
>>> clf_erm.fit(X=X_train, y=y_train, end_time=T)

>>> clf_ermlr.fit(X=X_train, y=y_train, end_time=T)
\end{CodeInput}
\end{CodeChunk}

Once this step is completed, the user gains access to a variety of useful methods. For instance, predictive probabilities can be obtained by calling the \texttt{predict\_proba()} method. 

\begin{CodeChunk}
\begin{CodeInput}
>>> clf_erm.predict_proba(X=X_test, end_time=T)

>>> clf_ermlr.predict_proba(X=X_test, end_time=T)
\end{CodeInput}
\end{CodeChunk}

Additionally, one can also get the prediction of the classifier by calling the \texttt{predict()}. 

\begin{CodeChunk}
\begin{CodeInput}
>>> clf_erm.predict(X=X_test, end_time=T)

>>> clf_ermlr.predict(X=X_test, end_time=T)
\end{CodeInput}
\end{CodeChunk}

Furthermore, the classifier's performance relative to the provided test labels can be evaluated by calling the \texttt{score()} method.

\begin{CodeChunk}
\begin{CodeInput}
>>> clf_erm.score(X=X_test, y=y_test, end_time=T)

>>> clf_ermlr.score(X=X_test, y=y_test, end_time=T)
\end{CodeInput}

\begin{CodeOutput}
0.65

0.88
\end{CodeOutput}
\end{CodeChunk}

Notably, it can be seen that classifier \textsc{ERMLR} performs better than classifier \textsc{ERM}, validating the theoretical results.

\begin{figure}[ht]
    \centering

    \begin{subfigure}[b]{0.49\textwidth}
        \centering
        \includegraphics[width=\linewidth]{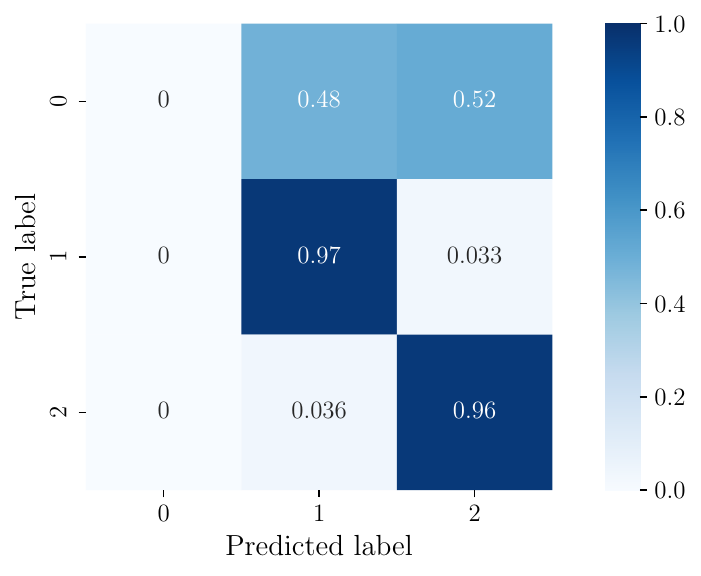}  
        \caption{ERM}
        \label{fig:left}
    \end{subfigure}
    \hfill
    \begin{subfigure}[b]{0.49\textwidth}
        \centering
        \includegraphics[width=\linewidth]{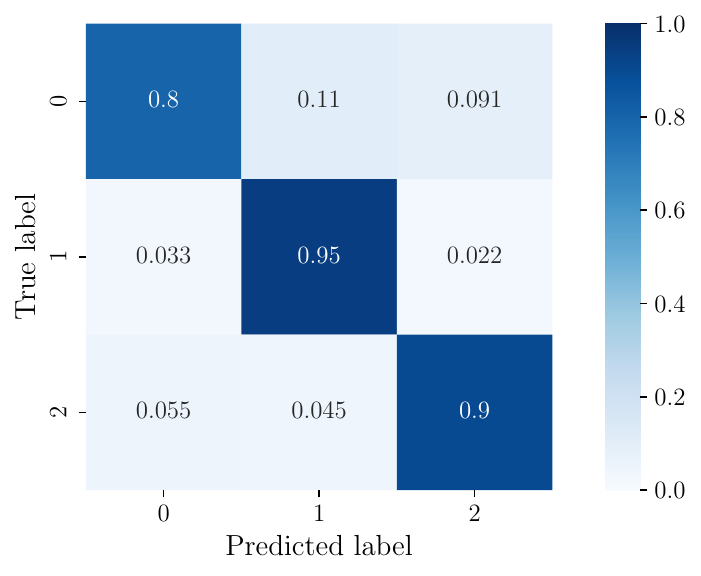}  
        \caption{ERMLR}
        \label{fig:right}
    \end{subfigure}

    \caption{Confusion matrices illustrating performance for the two classifiers. The diagonal shows the accuracy per class, while off-diagonal elements indicate misclassified instances.}
    \label{fig:E4_cm_ERM_ERMLR}
\end{figure}

Finally, the user can also plot the confusion matrix of the prediction with respect to the true labels by calling the \texttt{plot\_score\_cm()} method. The two confusion matrices associated with each classifier are presented in Figure~\ref{fig:E4_cm_ERM_ERMLR}.

\begin{CodeChunk}
\begin{CodeInput}
>>> clf_erm.plot_score_cm(X=X_test, y=Y_test, end_time=T)

>>> clf_ermlr.plot_score_cm(X=X_test, y=Y_test, end_time=T)
\end{CodeInput}
\end{CodeChunk}

\section{Illustration on MemeTracker dataset} \label{sec:illustrations}

In this section, we illustrate the \pkg{Sparklen} package using the MemeTracker dataset (see \cite{Leskovec:2009}). MemeTracker is a benchmark corpus containing blog posts published between August 2008 and April 2009. This dataset tracks frequently quoted phrases and memes, allowing for the analysis of information propagation across digital media. It is widely used in computational social science, natural language processing, and network analysis to study how narratives emerge, evolve, and spread across the internet.

\paragraph{Pre-processing.} We use posts from August 2008 to December 2008, covering a period of 153 days. An event for a website is defined as the creation of a post on that website containing a hyperlink to any other website. Additionally, we consider that an edge exists between two websites if at least one hyperlink connects them in the training set. To focus on the most influential sources, we extract the top 100 media sites with the highest number of documents. For each of these 100 websites, we retrieve their corresponding events by using the code made available by \cite{Achab:2017}. Finally, we segment the process by grouping events by day, resulting in 153 repetitions of events over the given period. 
Finally, it is important to note that an approximation of the interaction matrix for the network can be computed by leveraging hyperlinks to trace the flow of information (see \cite{Achab:2017}). These causal relationships relies on the parenthood relations given by the clustering representation (see \cite{Bacry:2015}). 

\paragraph{Benchmark.} For comparison purposes, we consider a benchmark approach, specifically \textsc{ADM4} (see \cite{Zhou:2013}). Both our procedure and \textsc{ADM4} operate with a given fixed input $\beta$. Since this hyperparameter is unknown, we explore a grid of decay values and select the one that maximizes the score, \textit{i.e.}, the loss value under consideration. Given the sparse nature of the ground truth interaction matrix, both procedures are coupled with Lasso regularization. For the \pkg{Sparklen} procedure, we use the least-squares loss for its ability to effectively handle a large number of events and the \textsc{EBIC} criterion to select the penalty constant. In the case of \textsc{ADM4}, no automatic tuning of the penalty constant is provided, so we explore a grid of regularization constants and select the one that maximizes the score.

\paragraph{Metrics.} Let $\alpha^*$ be the ground truth interaction matrix and $\hat{\alpha}$ an estimated candidate. We consider the following three metrics. First, the Hamming Distance: 
$$\text{HammDist}(\alpha^*, \hat{\alpha}) := \frac{1}{d^2}\sum_{j,j'=1}^d \one_{\{\alpha^*_{j,j'} \neq \hat{\alpha}_{j,j'}\}}$$
which allows to assess the quality of the support recovery. Then, the Relative Error:
$$\text{RelErr}(\alpha^*, \hat{\alpha}) := \frac{1}{d^2}\sum_{j,j'=1}^d \frac{\lvert \alpha^*_{j,j'} - \hat{\alpha}_{j,j'} \rvert}{\lvert \alpha^*_{j,j'} \rvert} \one_{\left\{ \alpha^*_{j,j'} \neq 0\right\}} + \lvert \hat{\alpha}_{j,j'} \rvert \one_{\alpha^*_{j,j'} = 0\}} $$
which quantifies the averaged relative error between the true $\alpha^*$ and the estimated $\hat{\alpha}$ on non-zero entries of $\alpha^*$ and error on zeros entries. Finally, the mean Kendall rank Correlation:
$$\text{RankCorr}(\alpha^*, \hat{\alpha}) := \frac{1}{d}\sum_{j=1}^d \tau(\alpha^*_{j, \cdot}, \hat{\alpha}_{j,\cdot})$$
which corresponds to the averaged Kendall’s rank correlation coefficient between each row of the true $\alpha^*$ and that of the estimated $\hat{\alpha}$.
While relevant, we wish to highlight that these three metrics have limitations in terms of reliability when used to compare against a sparse matrix, as is the case for the ground truth here. For example, the null matrix has a Hamming Distance and a Relative Error of $0.06$ compared to the ground truth matrix.

\paragraph{Evaluation scheme.} We present the evaluation scheme that relies on Monte-Carlo repetitions. We independently repeat the following steps 5 times. Following the approach in \cite{Zhou:2013}, the dataset is split evenly, with 50\% allocated for training and the remaining 50\% for testing. On the training data, we run both procedures by iterating through a grid of beta values and retrieve the estimate $\hat{\alpha}$  associated with the optimal decay parameter. Once this is done, we compute the metrics of the estimate provided by the two procedures. Finally, on the test dataset, we compute the negative log-likelihood of the model based on the estimated parameters.

\paragraph{Results.} The obtained results are given in Table~\ref{tab:results_meme}. 

\begin{table}[ht]
\centering
\begin{tabular}{llllll}
\toprule
\textbf{Metric} & \textbf{\text{HammDist}}  & \textbf{\text{RelErr}} &  \textbf{\text{RankCorr}} & \textbf{\text{NegLikTest}} \\ \midrule
\textbf{\textsc{ADM4}} & 0.80 (0.01) & 75.56 (2.60) & 0.136 (0.003) & \textbf{1.94e+07} (4.32e+05)\\
\textbf{Sparklen} & \textbf{0.06} (0.00) & \textbf{37.68} (0.20) & \textbf{0.570} (0.008) & 2.20e+07 (5.42e+05)\\
\bottomrule
\end{tabular}
\caption{Averaged values of the metrics over the 5 repetitions for both procedures, with the standard deviation provided in parentheses.}
\label{tab:results_meme}
\end{table}

As observed, our procedure yields a higher negative log-likelihood on the test sample compared to \textsc{ADM4}. This can be attributed to the fact that our method optimizes the least-squares loss, whereas \textsc{ADM4} is based on log-likelihood. On the other hand, it can be seen that our procedure outperforms \textsc{ADM4} across all three metrics. In particular, the estimated interaction matrix exhibits a very low Hamming Distance, suggesting that our method effectively recovers the support of the ground truth $\alpha^*$. This indicates that it captures the influence network more accurately.

\section{Concluding remarks and future perspectives} \label{sec:conclusion}

\paragraph{Conclusion.} \pkg{Sparklen} provides a collection of statistical learning tools designed to work with MHP in \proglang{Python}. In particular, it implements an efficient sampling procedure, a highly versatile and flexible framework for performing inference, and, finally, novel approaches to address the challenge of multiclass classification within the supervised learning framework. In this article, we have presented these tools and their use cases through instructive examples. Notably, we explored the potential and effectiveness of the proposed procedures using simulated data. Additionally, we demonstrated the inference framework by recovering a social network graph from the MemeTracker dataset. Together with its statistical contributions, its quality implementation, which ensures both accessibility and efficiency, makes it a highly appealing software solution. We strongly believe that this package can greatly benefit users seeking to employ MHP as a modeling tool across various applications.

\paragraph{Perspectives.} At the heart of this project is the ambition to provide the community with a cohesive environment of cutting-edge, user-friendly tools for manipulating MHP. While this package already establishes a solid foundation, plans for future developments are envisioned to further enhance and expand its scope of usability and accessibility. Let us discuss few of them. 

A natural first step in this incremental goal would be to incorporate additional kernel functions. While exponential kernels are widely used and cover a broad range of applications, introducing more complex options, such as sums of exponential kernels (see \cite{Lemonnier:2014}), would further expand the scope of applicability. From a implementation standpoint, this addition would present minimal challenges, as the library is designed in a modular fashion, enabling model classes to be seamlessly integrated into the inference machinery. The only requirement would be to implement the actual model classes, thus encouraging such an addition.

This article focuses on linear Hawkes processes, but incorporating tools to work with non-linear ones could be both interesting and highly motivating from an application standpoint. Allowing inhibition-type interactions would enhance modeling possibilities and extend applicability to a broader range of cases. For example, \cite{Bonnet:2023} introduces a method to estimate MHP processes with an exponential kernel in the presence of both excitation and inhibition effects.

Finally, incorporating a multi-threading option to leverage the computational capabilities of modern processors could significantly enhance efficiency in computationally demanding tasks.

\section*{Computational details}

The results in this paper were obtained using \proglang{Python}~3.11.10 with the \pkg{Sparklen}~0.1.0 package on a CentOS Linux release 7.7.1908 operating system. The computations were run on a single node with an Intel Xeon Gold 6230 CPU using one core and 16 GB of RAM.

\section*{Reproducibility and documentation}
All results and figures can be reproduced by running the scripts 
available in the GitHub repository \url{https://github.com/romain-e-lacoste/sparklen}. The repository includes all necessary code, along with instructions for setting up the required environment and dependencies. Additionally, to facilitate the use of \pkg{Sparklen}, the code is meticulously commented. Full documentation, taking advantage of these comments with the \proglang{Python} tool \pkg{Sphinx} (\cite{Brandl:2025}), is currently under construction. The \pkg{Sparklen} package is distributed under the OSI-approved 3-Clause BSD License.

\newpage
\section*{Acknowledgments}
This work has been supported by the Chaire \q{Modélisation Mathématique et Biodiversité} of Veolia-École polytechnique-Museum national d’Histoire naturelle-Fondation X, through a Ph.D. scholarship. The computations have been performed under the project \q{hawkes} on the HPC facility Cholesky operated by the IDCS / École polytechnique. This work is also part of the 2022 DAE 103 EMERGENCE(S) - PROCECO project supported by Ville de Paris. Finally, I am grateful to my advisors, Christophe Denis, Charlotte Dion-Blanc, and Laure Sansonnet, for their sound advice and their review work throughout the writing of this article. 

\bibliography{refs}

\newpage

\begin{appendix}

\section{Details on pre-calculation}
\label{app:precompute}

To avoid overloading the notations, we consider the version of functional that is not averaged over multiple repetitions but rather calculated from a single trajectory.

\paragraph{Log-likelihood.} The negative log-likelihood loss can be written as follows:
$$L_{T}(\theta) = \frac{1}{T}\sum_{j=1}^d L_{j, T}(\theta)$$.

Developing, for any $j \in [d]$, we have:

\begin{align*}
    L_{j, T}(\theta) &= \int_0^T \left(\mu_j + \sum_{j'=1}^d \sum_{h:t_{j',h}<t} \alpha_{j,j'}\beta e^{-\beta(t-t_{j',h})}\right) \dd t \\
    &\quad - \sum_{\ell:t_{j,\ell} < T}\log\left(\mu_j + \sum_{j'=1}^d \sum_{h:t_{j',h} < t_{j,\ell}}\alpha_{j,j'}\beta e^{-\beta(t_{j,\ell}-t_{j',h})}\right) \\
    &= \mu_jT + \sum_{j'=1}^d\alpha_{j,j'}\sum_{h:t_{j',h}<T}\left(1-e^{-\beta(T-t_{j',h})}\right) \\
    &\quad - \sum_{\ell:t_{j,\ell} < T}\log\left(\mu_j + \sum_{j'=1}^d \alpha_{j,j'}\sum_{h:t_{j',h} < t_{j,\ell}}\beta e^{-\beta(t_{j,\ell}-t_{j',h})}\right) \\
    &= \mu_jT + \sum_{j'=1}^d \mathcal{I}_{j'} - \sum_{\ell:t_{j,\ell} < T}\log\left(\mu_j + \sum_{j'=1}^d \alpha_{j,j'} \Psi_{j'}\left(t_{j,\ell}\right)\right)
\end{align*}

For the gradient, it follows that for $j \in [d]$, we have:
$$\frac{\partial L_{j, T}(\theta)}{\partial \mu_j} = T - \sum_{\ell:t_{j,\ell} < T} \frac{1}{\mu_j + \sum_{j'=1}^d \alpha_{j,j'} \Psi_{j'}\left(t_{j,\ell}\right)}.$$

In addition, for \( j, j' \in [d] \), we have:
$$\frac{\partial L_{j, T}(\theta)}{\partial \alpha_{j,j'}} = \mathcal{I}_{j'} - \sum_{\ell:t_{j,\ell} < T} \frac{\Psi_{j'}\left(t_{j,\ell}\right)}{\mu_j + \sum_{j'=1}^d \alpha_{j,j'} \Psi_{j'}\left(t_{j,\ell}\right)}.$$

\paragraph{Least-squares.} The least-squares loss can be written as follows:
$$R_{T}(\theta) = \frac{1}{T}\sum_{j=1}^d R_{j, T}(\theta)$$.

Developing, for any $j \in [d]$, we have:
\begin{align*}
    R_{j, T}(\theta) &= \int_0^T \left(\mu_j + \sum_{j'=1}^d \sum_{h:t_{j',h}<t} \alpha_{j,j'}\beta e^{-\beta(t-t_{j',h})}\right)^2 \dd t \\
    &\quad - 2 \sum_{\ell:t_{j,\ell} < T}\left(\mu_j + \sum_{j'=1}^d \sum_{h:t_{j',h} < t_{j,\ell}}\alpha_{j,j'}\beta e^{-\beta(t_{j,\ell}-t_{j',h})}\right) \\
    &= \mu_j^2 + \sum_{j'=1}^d\alpha_{j,j'}^2 \sum_{h:t_{j',h}<T} \frac{\beta}{2}\left(1-e^{-2\beta(T-t_{j',h})}\right) \\
    &\quad + \sum_{j'=1}^d\sum_{j''=1}^d\alpha_{j,j'}\alpha_{j,j''}\sum_{h:t_{j',h}<T}\left(1-e^{-2\beta(T-t_{j',h})}\right)\sum_{h':t_{j'',h'}<t_{j',h}}\beta e^{-\beta(t_{j',h}-t_{j'',h'})} \\
    &\quad + 2\mu_j\sum_{j'=1}^d\alpha_{j,j'}\sum_{h:t_{j',h}<T}\left(1-e^{-\beta(T-t_{j',h})}\right) \\
    &\quad - 2\mu_j N_j(T) - 2\sum_{j'=1}^d\alpha_{j,j'}\sum_{\ell:t_{j,\ell}<T}\sum_{h:t_{j',h}<t_{j,\ell}}\beta e^{-\beta(t_{j,\ell}-t_{j',h})} \\
    &= \mu_j^2 + \sum_{j'=1}^d\alpha_{j,j'}^2 \mathcal{I}_{j'}^2 + \sum_{j'=1}^d\sum_{j''=1}^d\alpha_{j,j'}\alpha_{j,j''} \mathcal{W}_{j', j''} \\
    &\quad + 2\mu_j\sum_{j'=1}^d\alpha_{j,j'}\mathcal{I}_{j'} -2\mu_j N_j(T) - 2\sum_{j'=1}^d\alpha_{j,j'} \mathcal{V}_{j, j'}
\end{align*}

For the gradient, it follows that for $j \in [d]$, we have:
$$\frac{\partial R_{j, T}(\theta)}{\partial \mu_j} = 2\mu_jT + 2\sum_{j'=1}^d\alpha_{j,j'}H_{j'}-2N_j(T)$$
In addition, for \( j, j' \in [d] \), we have:
$$\frac{\partial R_{j, T}(\theta)}{\partial \alpha_{j,j'}} = 2\alpha_{j,j'}\mathcal{I}^2_{j'} + \sum_{\substack{j"=1 \\ j''  \neq j'}}^d \alpha_{j,j''} \mathcal{W}_{j',j''} +2\mu_j\mathcal{I}_{j'} - 2\mathcal{V}_{j,j'}$$

Notably, such factorization unveils a remarkable computational property of the least-squares functional. Indeed, pre-calculation performed once in advance removes the complexity’s dependency on the number of events for both loss and gradient evaluation. This makes the loss function highly advantageous in studies involving networks with a very large number of observed events.

\section{Details on SWIG}
\label{app:mecanism}

\paragraph{\proglang{C++}/\proglang{Python} binding architecture.} To make the \proglang{C++} class usable from \proglang{Python}, it is wrapped in \proglang{Python} using \pkg{SWIG}. The proxy classes generated automatically by \pkg{SWIG} are placed in a build folder to indicate that, although they can be imported into \proglang{Python}, they are not intended for direct user interaction. Instead, a corresponding \proglang{Python} class is created to act as a shell around these \pkg{SWIG} proxy classes. This \proglang{Python} class contains an attribute referencing the proxy class, with each method being rewritten to include error management and to call the proxy method through the attribute for computation. This three-layer design allows error handling to be implemented and the classes to be given a fully \proglang{Python} interface leveraging \pkg{SWIG} generated code while benefiting from the high-level performance of \proglang{C++}.

\paragraph{Internal interoperability mechanism.} The integration between \proglang{C++} and \proglang{Python} in the \pkg{Sparklen} package is underpinned by a sophisticated mechanism designed to enable seamless data sharing between the two languages. At its heart lies the \texttt{SharedArray} class, a highly adaptable array container in \proglang{C++} that leverages the \texttt{std::shared\_ptr} smart pointer for efficient memory management. This implementation ensures automatic reference tracking, safeguarding against memory leaks. 

Then, this interoperability is made possible by the utilization of \pkg{SWIG} typemaps, specifically \texttt{typemap(in)} and \texttt{typemap(out)}. These typemaps are crafted to facilitate data conversion without incurring additional memory allocation, while carefully preserving ownership semantics. For instance, when a \proglang{Python} \texttt{NumPy} array is passed to \proglang{C++}, the \proglang{C++} \texttt{SharedArray} refrains from assuming ownership, instead accessing the memory directly via a pointer. This approach eliminates redundant copying and maximizes efficiency since no new memory allocation is done. 

Consider the following example, which shows how a 2D \texttt{NumPy} array is converted into a \texttt{SharedArrayDouble2D} object:

\begin{verbatim}
%typemap(in) (SharedArrayDouble2D &) (SharedArrayDouble2D res) {
    if (!PyArray_Check($input)) {
        PyErr_SetString(PyExc_TypeError, 
            "The input argument should be a NumPy Array");
        SWIG_fail;
    }
    if (PyArray_TYPE((PyArrayObject*)$input) != NPY_DOUBLE) {
        PyErr_SetString(PyExc_TypeError, 
            "The data type of the NumPy Array should be double");
        SWIG_fail;
    }
    if (PyArray_NDIM((PyArrayObject*)$input) != 2) {
        PyErr_SetString(PyExc_ValueError, 
            "The Expected Numpy Array should be 2-dimensional");
        SWIG_fail;
    }
    // Collect NumPy Array features
    npy_intp n_rows = PyArray_DIM((PyArrayObject*)$input, 0);
    npy_intp n_cols = PyArray_DIM((PyArrayObject*)$input, 1);
    double *data = static_cast<double*>(PyArray_DATA((PyArrayObject*)$input));

    // Instantiate SharedArray2D from NumPy array features
    res = SharedArrayDouble2D(n_rows, n_cols, data);
    res.setPythonOwner(true);
    $1 = &res;
}
\end{verbatim}

The \texttt{NumPy} array's dimensions and data pointer are extracted, enabling the creation of a \texttt{SharedArrayDouble2D} object directly linked to the same memory as the \texttt{NumPy} array. Ownership is carefully managed by flagging the \proglang{Python} ownership attribute, ensuring memory efficiency and conflict-free operations.

\end{appendix}


\end{document}